\documentclass[11pt,english]{article}
\pdfoutput=1
\usepackage{jheppub}
\usepackage{amsmath}
\usepackage{amssymb}
\usepackage{amsfonts}

\usepackage{graphicx}

\usepackage{braket}
\usepackage{wrapfig}
\usepackage{subcaption}

\newcommand{\beq}{\begin{equation}}
\newcommand{\eeq}{\end{equation}}

\newcommand{\ft}[2]{{\textstyle\frac{#1}{#2}}}
\newcommand{\fft}[2]{\frac{#1}{#2}}
\newcommand{\nn}{{\nonumber}}

\begin{document}

\preprint{MCTP-14-09}

\title{Hidden horizons in
non-relativistic AdS/CFT}

\author{Cynthia Keeler,}

\author{Gino Knodel}

\author{and James T. Liu}

\affiliation{Michigan Center for Theoretical Physics, Randall Laboratory
of Physics,\\
 The University of Michigan, Ann Arbor, MI 48109--1040, USA}

\emailAdd{keelerc@umich.edu} \emailAdd{gknodel@umich.edu} \emailAdd{jimliu@umich.edu}

\abstract{We study boundary Green's functions for spacetimes with non-relativistic scaling symmetry.  For this class of backgrounds, scalar modes with large transverse momentum, or equivalently low frequency, have an exponentially suppressed imprint on the boundary. We investigate the effect of these modes on holographic two-point functions. We find that the boundary Green's function is generically insensitive to horizon features on small transverse length scales. We explicitly demonstrate this insensitivity for Lifshitz $z=2$, and then use the WKB approximation to generalize our findings to Lifshitz $z>1$ and RG flows with a Lifshitz-like region. We also comment on the analogous situation in Schr\"odinger spacetimes.  Finally, we exhibit the analytic properties of the Green's function in these spacetimes.}

\maketitle


\section{Introduction}
One of the remarkable features of AdS/CFT is the notion that the degrees
of freedom in the bulk can be mapped onto a boundary theory in one
fewer dimension. After all, one ordinarily expects a scaling with
the dimensionality of the system, so that more degrees of freedom
could reside in the bulk than on the boundary. However, this conventional
picture breaks down for an AdS bulk, as points deep in the interior
can remain in causal contact with the boundary. Equivalently, radial
null geodesics in AdS can reach the boundary at a finite affine parameter.
Heuristically, this connection allows AdS/CFT to be viewed as both
a map from the bulk to the boundary and one from the boundary to the bulk.

Although the power of AdS/CFT has generally been in the ability to
investigate strongly coupled field theories by performing classical
computations in a weakly coupled bulk, there has  been considerable
interest in going from the boundary to the bulk as well \cite{Balasubramanian:1999ri,Banks:1998dd,Bousso:2012mh,Heemskerk:2012mn,Kabat:2013wga,Leichenauer:2013kaa,Keeler:2013msa,Rey:2014dpa}. One way this
has been addressed is through the notion of a smearing function that
maps boundary operators to bulk operators \cite{Bena:1999jv,Hamilton:2005ju,Hamilton:2006az}. For a pure AdS
bulk, the smearing function allows the reconstruction of any local
bulk operator in terms of local operators smeared along the
boundary. The existence of this smearing function provides support
to the idea that AdS/CFT is a true duality between bulk and boundary
theories. However, it turns out that the AdS case is rather special
in that generically only relativistic domain-wall flows seem to admit
a smearing function that reconstructs the entire bulk \cite{Leichenauer:2013kaa,Keeler:2013msa,Rey:2014dpa}.
In particular, the boundary to bulk mapping via smearing functions
breaks down for certain ``non-relativistic'' spacetimes, in which
Lorentz invariance along the transverse directions is broken \cite{Keeler:2013msa}.
Such spacetimes play an essential role in the holographic correspondence
between condensed matter systems and gravity (AdS/CMT) \cite{Kachru:2008yh,Charmousis:2010zz,Huijse:2011ef,Son:2008ye,Balasubramanian:2008dm,Adams:2008wt,Hartnoll:2009sz}.
Hence it is important to explore the implications of this breakdown in the smearing function
and to see what consequences there may be for non-relativistic AdS/CFT.

Heuristically, the breakdown of the smearing function for non-relativistic holography can
be traced to the fact that null geodesics carrying transverse momentum no longer reach the
boundary.  This leads to a decoupling of the deep IR from the boundary theory, at least when
using probes with large transverse momentum.  More precisely, for a spacetime with
boundary translational invariance, the equations of motion for a bulk field can be formulated as a
second-order radial equation by working in momentum space $(\omega,\vec k)$ of the boundary
theory.  The radial equation can then be transformed into an effective Schr\"odinger equation with
potential $U_{\rm eff}$ that depends on the precise form of the bulk metric.  It is then possible to
see that $U_{\rm eff}$ develops a tunneling barrier with height set by the transverse momentum
$|\vec k|$ whenever transverse Lorentz invariance is explicitly broken.  What this means is that
modes with large momentum are suppressed by an exponential
factor $e^{-c|\vec{k}|}$ (for some fixed constant $c$) when they reach the boundary. To reconstruct
the original amplitude of the signal, the observer has to multiply
his measured amplitude by an exponentially large number $e^{+c|\vec{k}|}$.
This leads to an exponential divergence and hence breakdown of the smearing function at
large $|\vec{k}|$.

While at some level the conventional approach to AdS/CFT as a mapping from the bulk to
the boundary is unaffected by the non-existence of a smearing function, the tunneling barrier
in the effective Schr\"odinger potential nevertheless leads to a decoupling of some of the
deep IR information from the boundary.  This decoupling has a direct physical consequence in terms of the
boundary Green's function $G(\omega,\vec k)$ for various non-relativistic spacetimes, including
Lifshitz and Schr\"odinger geometries.  The boundary Green's function is one of the most basic
quantities in AdS/CFT, and is computed by solving the classical equations of motion with
appropriate boundary conditions at the horizon.  For example, in Minkowski AdS/CFT, the
retarded Green's function is obtained by taking infalling boundary conditions, while the
advanced Green's function would be obtained by taking outgoing boundary conditions.
In this sense, the Green's function appears to probe the entire bulk geometry all
the way from the boundary to the horizon.  However, in any regime where $U_{\rm eff}$
develops a large tunneling barrier, the IR geometry decouples, and the boundary
Green's function no longer carries any information about the horizon.  (More precisely,
the horizon information is exponentially suppressed.)  A consequence of this is that multiple
bulk geometries can give rise to identical boundary Green's functions up to exponentially
small terms, at least in this decoupling regime.

Although it may seem surprising that a single boundary theory could admit multiple
holographic duals, it is important to note that this is only the case for a restricted range of momenta.
For example, in the case of Lifshitz spacetimes with exponent $z$, a large tunneling barrier
only occurs for $\omega\ll |\vec k|^z$, which lies well under the dispersion relation $\omega
\sim |\vec k|^z$.  (In the relativistic ($z=1$) case, this corresponds to spacelike momentum,
and the radial wavefunction tunnels all the way to the horizon.)
Nevertheless, it does indicate that the boundary Green's function in this regime
will be exponentially insensitive to how the Lifshitz horizon is resolved.

The IR-insensitivity of the boundary two-point function is more than
just a mathematical curiosity. This can be seen by considering a change of boundary conditions from infalling to outgoing at the horizon.
The response of the Green's function to such a flip of boundary conditions is captured by the spectral function
\begin{equation}
\chi(\omega,\vec{k})=2\,\mbox{Im}\,G_{R}(\omega,\vec{k})=-i\left(G_{R}(\omega,\vec{k})-G_{A}(\omega,\vec{k})\right),
\label{eq:spectral}
\end{equation}
which measures the density of states in the dual theory. The IR-insensitivity of $G(\omega,\vec k)$ manifests itself as a general insensitivity to horizon boundary conditions, which in turn
can be used to derive universal features of the spectral function in the region
of momentum-space that represents the tunneling modes.  We will demonstrate
this explicitly for the case of scalar Green's functions for
Lorentz-violating RG flows. Using the WKB approximation, we show that
for any such flow, the spectral function has a universal exponential
tail $\chi(\omega,\vec k)\sim e^{-c|\vec k|}$ (for some constant $c$) at low frequencies
$\omega$ and large momenta $|\vec{k}|$.

The observation that the spectral function becomes highly suppressed in the low frequency limit
has been made in the context of non-zero temperatures, as well as for fermions and vector fields
\cite{Son:2002sd,Horowitz:2009ij,Basu:2009xf,Hartnoll:2011dm,Iizuka:2011hg,Alishahiha:2012nm,Hartnoll:2012rj,Hartnoll:2012wm,Gursoy:2012ie}.  In particular, introducing a black hole into the bulk naturally generates a tunneling barrier, leading to exponential insensitivity to the horizon boundary conditions.  Our results demonstrate that this exponential suppression persists at zero temperature, and moreover occurs for any Lorentz-violating flow in the small $\omega$, large $|\vec k|$ limit.

This paper is organized as follows. In section~\ref{sec:greensf},
we review the holographic calculation of field theory Green's functions.
Using the WKB approximation, we show precisely how tunneling leads
to an exponential insensitivity of $G(\omega,\vec k)$ to horizon boundary conditions,
or equivalently to an exponentially small spectral function. In section~\ref{sec:Lifz2}
we present the analytic calculation of the Green's
function for $z=2$ Lifshitz. We demonstrate explicitly that tunneling
modes with $\omega\ll|\vec{k}|^{z}$ leave an exponentially small imprint
on the spectral function, and contrast this with the AdS case, where
modes with spacelike momenta are not part of the spectrum. We present
numerical results for other values of $z$ to show that this behavior
is not limited to the $z=2$ case, but is in fact a generic property
of Lifshitz spacetimes. In section~\ref{sec:rgflows}, we use WKB
methods to study the features of spectral functions for RG flows that
involve a Lifshitz-scaling region. For the specific example of a flow
from Lifshitz to $\mathrm{AdS_{2}}\times\mathbb{R}^{d}$, we show
that the low-energy behavior of the spectral function is determined
by IR physics. However, at large momenta $|\vec{k}|$, the numerical
coefficient between $\chi_{\mathrm{UV}}$ and $\chi_{\mathrm{IR}}$ is suppressed by
$\sim e^{-\mathrm{const.}\cdot|\vec{k}|}$ and it becomes ``exponentially
hard'' to probe IR physics, even at $\omega\rightarrow0$.
In section~\ref{sec:schroedinger},
we comment on the case of Schr\"odinger geometries. Using our previous
results, we can map the Schr\"odinger case to AdS or Lifshitz and thereby read
off the Green's functions.  In section~\ref{sec:analytic}, we analyze the analytic properties
of spectral
functions for AdS, Lifshitz spacetime with $z=2$, and Schr\"odinger spacetime with $z=2$ and $z=3/2$. Finally, in section \ref{sec:discussion}, we summarize and discuss the implications of our results.

\section{Horizon boundary conditions and the Green's function}
\label{sec:greensf}
In contrast with Euclidean AdS/CFT, in the Minkowski case, the Green's
function has a richer analytic structure that is closely related to
the causal propagation of information. For example, while the usual
computation of the retarded Green's function involves taking infalling
boundary conditions at the AdS horizon, one could equally well have
obtained the advanced Green's function by taking outgoing boundary
conditions. In the situation where time reversal invariance holds,
the retarded and advanced Green's functions are related by complex
conjugation. This is easy to understand in terms of boundary conditions
at the horizon, since complex conjugation of the radial wavefunction
interchanges infalling with outgoing boundary conditions.

More generally, the AdS/CFT Green's function probes the bulk, as its
computation depends on our ability to relate horizon with boundary
data. Consider, for example, the case of the scalar Green's function
arising from the action 
\begin{equation}
S=\int dt\, d^{d}x\, d\rho\sqrt{-g}\left[-\ft12\partial_{\mu}\phi\partial^{\mu}\phi-\ft12m^{2}\phi^{2}\right],\label{eq:scalaraction}
\end{equation}
in a background of the form
\begin{equation}
ds_{d+2}^{2}=e^{2A(\rho)}(-dt^{2}+d\rho^{2})+e^{2B(\rho)}d\vec{x}_{d}^{2}.\label{eq:lifgmet}
\end{equation}
The bulk solution takes the form 
\begin{equation}
\phi(t,\vec{x},\rho)=e^{i(\vec{k}\cdot\vec{x}-\omega t)}f_{\omega,\vec{k}}(\rho).
\end{equation}
For metrics of the form (\ref{eq:lifgmet}), the Klein-Gordon equation
$(\square-m^{2})\phi=0$ can be converted into a Schr\"odinger-like
equation 
\begin{equation}
-\psi''(\rho)+U(\rho)\psi(\rho)=\omega^{2}\psi(\rho),\label{eq:schr}
\end{equation}
where the effective potential is 
\begin{equation}
U=e^{2A}m^{2}+e^{2A-2B}\vec{k}^{2}+\left(\fft{dB}2\right)'^{2}+\left(\fft{dB}2\right)'',\label{eq:effpot}
\end{equation}
and where $f_{\omega,\vec{k}}(\rho)=e^{-dB/2}\psi(\rho)$. The reason
for our choice of gauge in the metric (\ref{eq:lifgmet}) is that
it directly leads to $\omega^{2}$ as an effective energy term in
the Schr\"odinger equation. Since the solution to the wave equation
will depend on both the bulk geometry and the horizon boundary condition,
the Green's function will similarly depend on the bulk and horizon
data. 

Now, let us assume that the metric is asymptotically of the Lifshitz (non-relativistic scale-invariant)
form 
\begin{equation}
ds_{d+2}^{2}\sim\left(\fft{L}{z\rho}\right)^{2}(-dt^{2}+d\rho^{2})+\left(\fft{L}{z\rho}\right)^{2/z}d\vec{x}_{d}^{2}.\label{eq:metLif}
\end{equation}
In this case, the asymptotic boundary solution to (\ref{eq:schr})
has the form 
\begin{equation}
\psi(\rho\to0)\sim A\left(\fft{z\rho}L\right)^{\fft12-\nu_{z}}+B\left(\fft{z\rho}L\right)^{\fft12+\nu_{z}},\label{eq:psiasympt}
\end{equation}
where 
\begin{equation}
\nu_{z}=\fft1z\sqrt{(mL)^{2}+\left(\fft{d+z}2\right)^{2}}.
\end{equation}
The holographic prescription for calculating boundary Green's functions
of $\phi$ is \cite{Son:2002sd}
\begin{equation}
G(\omega,\vec{k})=K\fft{B}A,
\end{equation}
where K is a numerical normalization constant. This result simply states
that the AdS/CFT Green's function is proportional to the ratio of
the normalizable to the non-normalizable mode.

The coefficients $B$ and $A$ are determined by solving the equation
(\ref{eq:schr}) subject to infalling or other appropriate boundary
conditions at the horizon. Assuming $U(\rho)$ approaches a constant
value $U_{0}$ at the horizon, the horizon solution has the form 
\begin{equation}
\psi\sim ae^{i\sqrt{\omega^{2}-U_{0}}\rho}+be^{-i\sqrt{\omega^{2}-U_{0}}\rho},\label{eq:horizab}
\end{equation}
and is oscillatory in the classically allowed range of frequencies,
$\omega^{2}> U_{0}$. The $a$ mode is infalling, while the $b$
mode is outgoing for positive $\omega$. In the forbidden range, we
may take $\sqrt{\omega^{2}-U_{0}}\to i\sqrt{U_{0}-\omega^{2}}$, so
the $a$ mode is exponentially damped, while the $b$ mode blows up.
Although the retarded Green's function is obtained by taking $b=0$,
here we leave it arbitrary so that we can examine the effect of changing the horizon boundary conditions.

Since the wave equation is second order and linear, the horizon and
boundary data are related by a linear transformation
\begin{equation}
\begin{pmatrix}A\\
B
\end{pmatrix}=\mathcal{M}\begin{pmatrix}a\\
b
\end{pmatrix}=\begin{pmatrix}\mathcal{M}_{Aa} & \mathcal{M}_{Ab}\\
\mathcal{M}_{Ba} & \mathcal{M}_{Bb}
\end{pmatrix}\begin{pmatrix}a\\
b
\end{pmatrix},
\end{equation}
where the connection matrix $\mathcal{M}$ depends on the bulk geometry
connecting the horizon to the boundary through the effective potential
(\ref{eq:effpot}). In terms of this matrix $\mathcal{M}$, the Green's
function then has the form
\begin{equation}
G(\omega,\vec{k})=K\fft{\mathcal{M}_{Ba}+\mathcal{M}_{Bb}(b/a)}{\mathcal{M}_{Aa}+\mathcal{M}_{Ab}(b/a)}.\label{eq:Ggen}
\end{equation}
This explicitly demonstrates how the Green's function connects the horizon (represented by
the horizon data $b/a$) to the boundary via the bulk matrix $\mathcal{M}$. 
We can, in fact, say a bit more about the
matrix $\mathcal{M}$. Since we are solving a real differential equation
(\ref{eq:schr}), any time $\psi$ is a solution, so is its complex
conjugate $\psi^{*}$. This allows us to relate the $a$ and $b$
modes in (\ref{eq:horizab}) whenever the solution is oscillatory
at the horizon. In particular, $\mathcal{M}_{Ab}=\mathcal{M}_{Aa}^{*}$,
and likewise $\mathcal{M}_{Bb}=\mathcal{M}_{Ba}^{*}$. In this case,
we obtain the expression
\begin{equation}
G(\omega,\vec{k})=K\fft{\mathcal{M}_{Ba}}{\mathcal{M}_{Aa}}\fft{1+e^{-2i\arg\mathcal{M}_{Ba}}(b/a)}{1+e^{-2i\arg\mathcal{M}_{Aa}}(b/a)}.\label{eq:Gba}
\end{equation}
This expression highlights the dependence of the Green's function on the ratio $b/a$
specifying the boundary condition at the horizon. The retarded Green's
function is obtained by taking $b/a=0$, while the advanced Green's
function corresponds to $b/a\to\infty$. Since $\mathcal{M}_{Ba}e^{-2i\arg\mathcal{M}_{Ba}}=\mathcal{M}_{Ba}^{*}$
(and likewise for $\mathcal{M}_{Aa}$), we may explicitly see that
$G_{A}(\omega,\vec{k})=G_{R}(\omega,\vec{k})^{*}$.

More generally, the Green's function expression (\ref{eq:Gba}) allows
us to explore the sensitivity of the boundary behavior to small changes
in the infrared. For example, a small change to the bulk geometry
in the deep IR would induce a change to the effective potential $U$
near the horizon. As a result, an infalling wave could scatter off
the perturbation, so that at some distance outside the horizon (but
still in the IR), the actual solution is mostly infalling, but now
picks up a small outgoing component as well. In this case, the effect
of the perturbation on the retarded Green's function can be modeled
by taking $b/a$ small but non-vanishing, so that a small outgoing
component is introduced. Expanding to lowest order in $b/a$, the
result is
\begin{equation}
G(\omega,\vec{k})=K\fft{\mathcal{M}_{Ba}}{\mathcal{M}_{Aa}}\left[1+(e^{-2i\arg\mathcal{M}_{Ba}}-e^{-2i\arg\mathcal{M}_{Aa}})\left(\fft{b}a\right)+\cdots\right].
\end{equation}
For generic values of the arguments, the sensitivity of the Green's
function to $b/a$ is of $\mathcal{O}(1)$. However, it becomes completely
insensitive to $b/a$ (and not just to leading order) in the limit
$\arg\mathcal{M}_{Ba}=\arg\mathcal{M}_{Aa}$. Note that in this limit,
the Green's function is purely real, as the ratio $\mathcal{M}_{Ba}/\mathcal{M}_{Aa}$
is real. Equivalently, the spectral function, (\ref{eq:spectral}),
goes to zero. Throughout this paper, we will therefore take an exponentially small $\chi$
as a signal for the insensitivity to a change of the near-horizon bulk state and/or geometry.

\subsection{Tunneling barriers and decoupling of the IR}

As we have seen above, when $\arg\mathcal{M}_{Ba}=\arg\mathcal{M}_{Aa}$,
the Green's function becomes purely real and thus invariant under
changing from retarded (infalling) to advanced (outgoing) boundary
conditions. This is actually not surprising, as complex conjugation
of a real function leaves it unchanged. What may appear more unusual
is that in this case, since the dependence on $b/a$ completely drops
out, the Green's function is unaffected by any choice of horizon boundary
conditions $0\leq |b/a|\leq\infty$.

It is important to note, however, that since the second order wave
equation admits two linearly independent solutions, the connection
matrix $\mathcal{M}$ is necessarily invertible. What this means is
that $\arg\mathcal{M}_{Ba}$ can never actually be degenerate with
$\arg\mathcal{M}_{Aa}$. As a result, the Green's function is never
real (in the classically allowed range of $\omega$), although it
can approach a real function in the limiting case. In this sense,
the horizon boundary conditions never completely drop out of the Green's
function computation. However, the dependence on the horizon can become
highly suppressed whenever $\mathcal{M}$ becomes nearly degenerate.

Since the effective Schr\"odinger equation (\ref{eq:schr}) governing
the wavefunction is specified by the effective potential (\ref{eq:effpot}),
the connection matrix $\mathcal{M}$ will depend on the explicit form
of $U$ as well as the frequency $\omega$. Here it is important to
note that, while the boundary is in a classically forbidden region,
the asymptotic form of the potential $U\sim1/\rho^{2}$ is too steep
for tunneling. This is the reason we have power law behavior at the
boundary and not exponential. If the shape of the potential is such
that there is no tunneling between the horizon and the boundary, then
the entries in $\mathcal{M}$ are all of $\mathcal{O}(1)$, and generically
there is no degeneracy. In this case, the UV and IR are tied together
by an $\mathcal{O}(1)$ transformation, and perturbations in the IR
are directly reflected in changes to the Green's function.

On the other hand, if the potential $U$ admits a tunneling region
and $\omega$ is below the barrier, then
the connection matrix $\mathcal{M}$ will become nearly degenerate.
This is exactly the situation where the Green's function becomes insensitive
to the horizon boundary conditions. Heuristically, what is going on
is that the tunneling barrier decouples the IR from the UV, so information
at the horizon becomes hidden from the boundary.

We may use a WKB approximation (see appendix \ref{sec:WKB}) to make the connection between tunneling
of the wavefunction and the form of ${\cal M}$ more precise.  Assuming
asymptotically Lifshitz behavior, the potential $U$ behaves near the boundary
as
\begin{equation}
U\left(\rho\rightarrow0\right)\sim\frac{\nu^{2}-1/4}{\rho^{2}}.
\end{equation}
We assume that the effective Schr\"odinger energy $\omega^2$ in (\ref{eq:schr}) is
such that the horizon falls into a classically allowed region.  Since the potential
increases without bound as we move towards the boundary, we will always encounter
a classical turning point $\rho_0$.  The wavefunction is thus oscillating in the classically
allowed region $\rho>\rho_{0}$ (corresponding to the IR) and growing/decaying
in the forbidden region $\rho<\rho_{0}$%

\begin{equation}
\psi_{\mathrm{WKB}}\approx\begin{cases}
\sqrt{\nu}\left(U-\omega^{2}\right)^{-\frac{1}{4}}\left(Ce^{S\left(\rho,\rho_{0}\right)}+De^{-S\left(\rho,\rho_{0}\right)}\right), & \rho<\rho_{0};\\
\sqrt{\nu}\left(\omega^{2}-U\right)^{-\frac{1}{4}}\left(ae^{i\Phi\left(\rho_{0},\rho\right)}+be^{-i\Phi\left(\rho_{0},\rho\right)}\right), & \rho>\rho_{0}.
\end{cases}\label{eq:WKBansatz}
\end{equation}
Here we defined
\begin{equation}
S\left(\rho,\rho_{0}\right)\equiv\int_{\rho}^{\rho_{0}}d\rho\sqrt{U-\omega^{2}},\qquad\Phi\left(\rho_{0},\rho\right)\equiv\int_{\rho_{0}}^{\rho}d\rho\sqrt{\omega^{2}-U}.\label{eq:SPhi}
\end{equation}
The functions $S$ and $\Phi$ carry all the relevant information
about the potential. As highlighted in \cite{Keeler:2013msa}, the WKB
approximation is somewhat subtle for a $1/\rho^2$ potential.  However, it remains
valid, provided we perform the shift $\nu^{2}\rightarrow\nu^{2}+1/4$.

The coefficients in \eqref{eq:WKBansatz} are tied together via the connection formulae
\begin{equation}
\left(\begin{array}{c}
C\\
D
\end{array}\right)={\cal M}^{\prime\prime}\left(\begin{array}{c}
a\\
b
\end{array}\right)=\left(\begin{array}{cc}
e^{-i\frac{\pi}{4}} & e^{i\frac{\pi}{4}}\\
\frac{1}{2}e^{i\frac{\pi}{4}} & \frac{1}{2}e^{-i\frac{\pi}{4}}
\end{array}\right)\left(\begin{array}{c}
a\\
b
\end{array}\right).\label{eq:Mpp}
\end{equation}
To relate the WKB coefficients $C$ and $D$ to the coefficients $A$
and $B$ in \eqref{eq:psiasympt}, we match $\psi_{\mathrm{WKB}}$
with the exact solution \eqref{eq:psiasympt} at some UV cutoff $\rho=\epsilon$,
which will be taken to zero at the end. The result can be written
as another matrix equation:
\begin{equation}
\left(\begin{array}{c}
A\\
B
\end{array}\right)={\cal M}^{\prime}\left(\begin{array}{c}
C\\
D
\end{array}\right) =\left(\begin{array}{cc}
{\cal M}_{AC}^{\prime} & {\cal M}_{AD}^{\prime}\\
{\cal M}_{BC}^{\prime} & {\cal M}_{BD}^{\prime}
\end{array}\right)\left(\begin{array}{c}
C\\
D
\end{array}\right).
\end{equation}
Combining this with \eqref{eq:Mpp}, we then find that ${\cal M=M^{\prime}}{\cal M}^{\prime\prime}$,
which can be used to find the Green's function \eqref{eq:Ggen} in
the WKB approximation.  To determine ${\cal M}^{\prime}$ explicitly, let us write
\begin{align}
\psi_{\mathrm{exact}} & =A\phi_{1}+B\phi_{2},\nn\\
\psi_{\mathrm{WKB}} & =C\phi_{3}+D\phi_{4},
\end{align}
with $\phi_{1/2}$ being the exact solution with boundary behavior
$\phi_{1/2}\approx\rho^{\frac{1}{2}\mp\nu}$, and
\begin{equation}
\phi_{3/4}\equiv\sqrt{\nu}\left(U-\omega^{2}\right)^{-\frac{1}{4}}e^{\pm S\left(\rho,\rho_{0}\right)}.
\end{equation}
The matching matrix is then given by
\begin{equation}
{\cal M}^{\prime} =\fft1{W_{12}}\left(\begin{array}{cc}
{W_{32}} & {W_{42}}\\
{W_{13}} & {W_{14}}
\end{array}\right),\quad \mbox{where}\quad W_{ij}\equiv\phi_{i}\left(\epsilon\right)\phi_{j}^{\prime}\left(\epsilon\right)-\phi_{i}^{\prime}\left(\epsilon\right)\phi_{j}\left(\epsilon\right).
\end{equation}
Working near the boundary, this takes the explicit form
\begin{equation}
\mathcal M' =\left(\begin{array}{cc}
\epsilon^{\nu}e^{S\left(\epsilon,\rho_{0}\right)} & 0\\
0 & \epsilon^{-\nu}e^{-S\left(\epsilon,\rho_{0}\right)}
\end{array}\right).
\end{equation}
We can easily read off the imaginary part of the Green's function and find

\begin{equation}
2\,\mathrm{Im}\,G_{\mathrm{WKB}}(\omega,\vec{k})
=K\frac{{\cal M^{\prime}}_{BD}}{{\cal M^{\prime}}_{AC}}
\frac{1-\big|\frac{b}{a}\big|^{2}}{1+\big|\frac{b}{a}\big|^{2}}
=K\epsilon^{-2\nu}e^{-2S\left(\epsilon,\rho_{0}\right)}
\frac{1-\big|\frac{b}{a}\big|^{2}}{1+\big|\frac{b}{a}\big|^{2}}.\label{eq:chiWKB}
\end{equation}
In the case $b=0$, corresponding to infalling conditions at the horizon, the above expression is simply the spectral function $\chi$. As we show in appendix \ref{sec:WKB}, the error due to the WKB approximation can be kept under perturbative control.
The dependence on the shape of the effective potential $U$ is
captured in the $e^{-2S}$ term in (\ref{eq:chiWKB}). While the near-boundary $1/\rho^{2}$
behavior only leads to power-law scaling, any tunneling region with
$U$ falling off slower than $1/\rho^{2}$ leads to an exponential
suppression factor in the spectral function.

More concretely, consider a spacetime that enjoys Lifshitz scaling
in some region in the bulk. The potential takes the form \eqref{eq:effpot},
with a tunneling term $\vec{k}^{2}e^{2(A-B)}\sim\vec{k}^{2}\rho^{2(1/z-1)}$.
Tunneling of the wavefunction through this part of the potential leads
to an exponential fall-off of the spectral function at large momenta
$|\vec{k}|$:
\begin{equation}
{\cal \chi}\left(\omega,\vec{k}\gg c^{-1}\right)=f\left(\omega\right)e^{-c|\vec{k}|},\label{eq:Asuppressed}
\end{equation}
with some geometry-dependent constant $c$. For some special cases
like pure Lifshitz, this constant can actually secretly carry an
additional dependence on $\vec{k}$ and $\omega$, making $\chi$ vanish
even faster. We will comment on this issue at the end of the next
section. From \eqref{eq:Asuppressed}, we see that changing from infalling
to outgoing boundary conditions results only in an exponentially small
change $\delta G\sim \chi \sim e^{-c|\vec{k}|}$. For $|\vec{k}|\rightarrow\infty$,
$\chi\rightarrow0$ and so the Green's function becomes purely
real, completely decoupling the near-horizon boundary conditions.
This establishes the insensitivity of the Green's function to IR physics.
We will further illustrate the connection between horizon boundary
conditions and IR physics in section \ref{sec:rgflows}.

\section{Horizon decoupling for Lifshitz backgrounds}

\label{sec:Lifz2}For general backgrounds, the connection matrix $\mathcal{M}$
and the resulting Green's function (\ref{eq:Ggen}) will have to be
obtained either numerically, or using approximation methods such as
WKB. However, analytic solutions are known for simple backgrounds
such as AdS and Lif$_{z=2}$. Here we highlight and contrast these
two cases as an explicit demonstration of the decoupling of the IR
in a Lifshitz background. In particular, we will confirm our prediction
\eqref{eq:Asuppressed} for the exponential fall-off of $\chi$ in
the Lifshitz case.

\subsection{The $z=1$ AdS case}
For a pure Lifshitz or AdS geometry, we can take the metric (\ref{eq:metLif})
to be exact throughout the bulk. In this case, the effective potential
becomes:
\begin{equation}\label{eq:LifU}
U=\fft{\nu_{z}^{2}-1/4}{\rho^{2}}+\vec{k}^{2}\left(\fft{L}{z\rho}\right)^{2-2/z}.
\end{equation}
Let us first consider the AdS case, which corresponds to $z=1$. Here
the potential is purely $1/\rho^{2}$ on top of a constant offset,
and there is no tunneling region (so long as $\omega\ge|\vec{k}\,|$).
The $1/\rho^{2}$ potential is ``too steep for tunneling'', and
the wavefunction grows or decays polynomially. The exact solution
for $\psi(\rho)$ is well known, and is given by a linear combination
of Bessel functions
\begin{equation}
\psi=\sqrt{\rho}\left[\alpha J_{\nu}(q\rho)+\beta Y_{\nu}(q\rho)\right],
\end{equation}
where $q=\sqrt{\omega^{2}-\vec{k}^{2}}=\sqrt{-k_{\mu}k^{\mu}}$. In
this case, it is straightforward to obtain
\begin{equation}
\mathcal{M}_{z=1}=\begin{pmatrix}\fft{\Gamma(\nu)}{\sqrt{\pi}}\left(\fft{qL}2\right)^{\fft12-\nu}e^{i(\fft\nu2-\fft14)\pi}\quad & \fft{\Gamma(\nu)}{\sqrt{\pi}}\left(\fft{qL}2\right)^{\fft12-\nu}e^{-i(\fft\nu2-\fft14)\pi}\\
\fft{\Gamma(-\nu)}{\sqrt{\pi}}\left(\fft{qL}2\right)^{\fft12+\nu}e^{-i(\fft\nu2+\fft14)\pi}\quad & \fft{\Gamma(-\nu)}{\sqrt{\pi}}\left(\fft{qL}2\right)^{\fft12+\nu}e^{i(\fft\nu2+\fft14)\pi}
\end{pmatrix},
\end{equation}
at least for non-integer values of $\nu$. Note that this has the
form
\begin{equation}
\mathcal{M}=\begin{pmatrix}\mathcal{M}(\nu)e^{i\varphi(\nu)} & \mathcal{M}(\nu)e^{-i\varphi(\nu)}\\
\mathcal{M}(-\nu)e^{i\varphi(-\nu)} & \mathcal{M}(-\nu)e^{-i\varphi(-\nu)}
\end{pmatrix},\label{eq:Mform}
\end{equation}
where $\varphi(\nu)=(\nu/2-1/4)\pi$. This form is related to the
$\nu\to-\nu$ symmetry of the effective potential.

For $\omega\ge|\vec{k}\,|$, the AdS Green's function can be obtained
from (\ref{eq:Gba}). Using relativistic notation, we find
\begin{equation}
G(q)=K\fft{\Gamma(-\nu)}{\Gamma(\nu)}\left(\fft{qL}2\right)^{2\nu}e^{-i\nu\pi}\fft{1+e^{i(\nu+\fft12)\pi}(b/a)}{1+e^{-i(\nu-\fft12)\pi}(b/a)}.\label{eq:AdSG}
\end{equation}
Recall that the retarded Green's function corresponds to taking $b/a=0$.
In order to examine the sensitivity to horizon boundary conditions,
we may expand to first order in $b/a$
\begin{equation}
G(q)=K\fft{\Gamma(-\nu)}{\Gamma(\nu)}\left(\fft{qL}2\right)^{2\nu}e^{-i\nu\pi}\left(1-2\sin(\nu\pi)\fft{b}a+\cdots\right).\label{eq:AdSGexp}
\end{equation}
Since we have assumed $\nu$ to be non-integral, this shows that $G(q)$
has $\mathcal{O}(1)$ sensitivity to the choice of horizon boundary
conditions $b/a$. Moreover, this sensitivity is present in both the
real and imaginary parts of the Green's function. For the spectral
function we find $\chi\sim q^{2\nu}$, as required by scale invariance, but no exponential suppression
factor.

\subsection{The $z=2$ Lifshitz case}

\begin{figure}
\begin{centering}
\hspace{-1cm}\includegraphics[height=7cm]{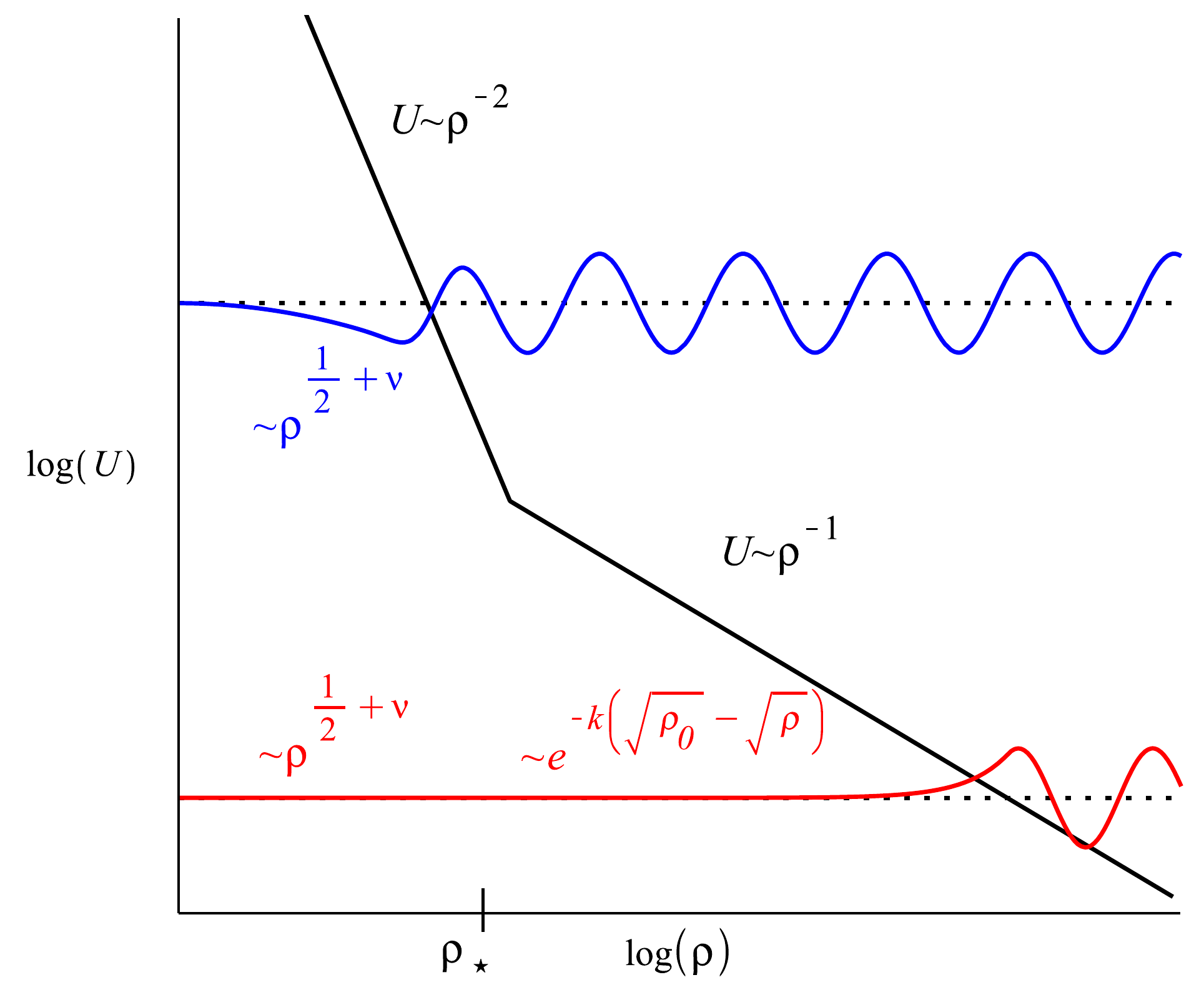}
\par\end{centering}
\caption{\label{fig:tunnsketc}Sketch of the effective potential $U$ for $z=2$
Lifshitz spacetime. The potential changes from the near-boundary $1/\rho^{2}$
behavior to the tunneling potential $U\sim1/\rho$ near the crossover
scale $\rho_{*}\sim1/|\vec{k}|^{2}$. A normalizable wavefunction
with large energy $\omega$ and low momenta $|\vec{k}|$ crosses the
barrier in the $1/\rho^{2}$ region and decays polynomially, according
to \eqref{eq:psiasympt} (blue curve). For low energies and large
momenta the crossing point lies within the tunneling region and the
wavefunction decays exponentially at first (red curve). This has the effect that states that are localized close to the horizon have an exponentially small amplitude at the boundary}
\end{figure}

We now turn to $z=2$ Lifshitz as an analytic example of a non-relativistic
system. Here the potential has a combination of $1/\rho^{2}$ and
$1/\rho$ terms
\begin{equation}
U_{z=2}=\fft{\nu_{2}^{2}-1/4}{\rho^{2}}+\fft{\vec{k}^{2}L/2}\rho.\label{eq:ULifz2}
\end{equation}
As is well known from quantum mechanics, the $1/\rho$ potential is
shallow enough that it presents a tunneling barrier in the system.
However, not all the modes have to tunnel through this part of the
potential. Denoting the crossover scale between $1/\rho$ and $1/\rho^{2}$
behavior as $\rho_{*}={2\nu_{2}^{2}}/{\vec{k}^{2}L}$, the condition
for a mode to tunnel is
\begin{equation}
\frac{\vec{k}^{2}L}{2\rho_{*}}\gg\omega^{2}\quad\Longrightarrow\quad
\alpha\equiv\frac{\vec{k}^{2}L}{2\omega}\gg\nu_{2}.
\end{equation}
For these modes, we expect an exponential suppression in $\alpha$, as
sketched in Figure~\ref{fig:tunnsketc}.

The potential (\ref{eq:ULifz2}) admits an analytic solution in terms
of the Whittaker functions $M_{-i\alpha/2,\nu_{2}}(-2i\omega\rho)$
and $W_{-i\alpha/2,\nu_{2}}(-2i\omega\rho)$. Since the potential
still has a $\nu_{2}\to-\nu_{2}$ symmetry, the connection matrix
maintains the form (\ref{eq:Mform}), however with
\begin{equation}
\mathcal{M}(\nu_{2})e^{i\varphi(\nu_{2})}=\Gamma(2\nu_{2})(\omega L)^{\fft12-\nu_{2}}e^{\pi\alpha/4}\fft{e^{i(\fft{\nu_{2}}2-\fft14)\pi}2^{i\alpha/2}}{\Gamma(\fft12+\nu_{2}+\fft{i\alpha}2)}.
\end{equation}
In contrast with the relativistic case, this function depends on the
ratio of $\vec{k}^{2}$ and $\omega$ through the parameter $\alpha$.
Using (\ref{eq:Gba}), the Green's function is then
\begin{equation}
G(\omega,\vec{k})=K\fft{\Gamma(-2\nu_{2})}{\Gamma(2\nu_{2})}(\omega L)^{2\nu_{2}}\fft{\Gamma(\fft12+\nu_{2}+\fft{i\alpha}2)}{\Gamma(\fft12-\nu_{2}+\fft{i\alpha}2)}e^{-i\nu_{2}\pi}\fft{1+e^{-2i\varphi(-\nu_{2})}(b/a)}{1+e^{-2i\varphi(\nu_{2})}(b/a)},\label{eq:GLif2}
\end{equation}
where
\begin{equation}
\varphi(\nu_{2})=\left(\fft{\nu_{2}}2-\fft14\right)\pi+\fft\alpha2\log2+\arg\Gamma(\ft12+\nu-\ft{i\alpha}2).
\end{equation}
For the non-tunneling modes with small $\alpha$, we find (to first
order in $b/a$)
\begin{eqnarray}
G(\omega,\vec{k}) & = & K\fft{\Gamma(-\nu_{2})}{\Gamma(\nu_{2})}\left(\fft{\omega L}4\right)^{2\nu_{2}}e^{-i\nu_{2}\pi}\left(1+\fft{i\pi\alpha}2\tan(\nu_{2}\pi)+\mathcal{O}(\alpha^{2})\right)\nn\\
 &  & \kern-1em \times\left[1-2\sin(\nu_{2}\pi)\left(1+\fft{i\alpha}2\left(i\pi-\log4+\psi(\ft12+\nu_{2})+\psi(\ft12-\nu_{2})\right)+\mathcal{O}(\alpha^{2})\right)\fft{b}a+\cdots\right],\nn\\
\end{eqnarray}
which matches the AdS Green's function (\ref{eq:AdSG}) in the limit
$\alpha\to0$ once we identify $L\to2L$, $\nu_{2}\to\nu$ and $\omega\to q$.
This should not be surprising because $\alpha\to0$ can be achieved
by taking $\vec{k}\to0$. In this limit the transverse space becomes
irrelevant, and the Lifshitz potential may be identified with the
AdS potential. As a result, the Green's function at small $\alpha$
is sensitive to the horizon boundary conditions in essentially the
same manner as given in (\ref{eq:AdSGexp}).

What is more interesting is the $\alpha\gg\nu_{2}$ limit, where the
horizon modes must tunnel under the $1/\rho$ potential to reach the
boundary. For large $\alpha$ we first use Stirling's approximation
to see that
\begin{equation}
\varphi(\nu_{2})\sim\fft\alpha2\left(1-\log\fft\alpha4\right)-\fft\pi4+\mathcal{O}\left(\fft1\alpha\right).
\end{equation}
A key observation is that, at leading order, the $\nu_{2}$ dependence
completely cancels out from the phase, and this is exactly what is
required for the Green's function (\ref{eq:GLif2}) to become insensitive
to the horizon boundary conditions. Beyond leading order, we may use
the identity
\begin{equation}
\varphi(\nu)-\xi(-\nu)=-\,\mathrm{Im}\,\log\left(1+e^{-2\pi i\nu-\pi\alpha}\right),
\end{equation}
obtained by application of the reflection formula $\Gamma(1-z)\Gamma(z)=\pi\csc(\pi z)$,
to see that $\varphi(\nu_{2})$ is an even function of $\nu_{2}$
to any finite order in the perturbative expansion in $1/\alpha$.
Explicitly, what we find is
\begin{eqnarray}
G(\omega,\vec{k})&=& K\fft{\Gamma(-2\nu_{2})}{\Gamma(2\nu_{2})}\left(\fft{|\vec{k}|L}2\right)^{4\nu_{2}}\left(1+e^{-i2\pi\nu_{2}}e^{-\pi\alpha}+\cdots\right)\nn\\
&&\kern10em\times
\left(1-2\sin(2\nu_{2}\pi)\left(\fft\alpha{4e}\right)^{i\alpha}e^{-\pi\alpha}\fft{b}a+\cdots\right).
\end{eqnarray}
This clearly demonstrates the insensitivity of the Green's function
to the horizon boundary conditions in the tunneling (large $\alpha$)
regime. It is important to note that the magnitude of the Green's
function is not necessarily small in this regime, and that it is only
the dependence on $b/a$ that is being exponentially suppressed.

The same conclusion can be drawn by looking at the spectral function:
\begin{equation}
\chi(\omega,\vec{k})= 2K\fft{\Gamma(-2\nu_{2})}{\Gamma(2\nu_{2})}\left(\fft{|\vec{k}|L}2\right)^{4\nu_{2}}\sin\left(-2\pi\nu_{2}\right)e^{-\pi\alpha}.
\end{equation}
At large $\alpha$, $\chi(\omega,\vec k)$ is exponentially small, as predicted in
the previous section. In the $\alpha\rightarrow\infty$ limit, the
spectral function vanishes and $G(\omega,\vec k)$ becomes completely insensitive
to changing boundary conditions.

One interesting aspect of pure Lifshitz spacetime is that the exponential
suppression is in the variable $\alpha\sim{\vec{k}^{2}}/\omega$, instead of just $|\vec{k}|$.
Again, the WKB approximation can help
us understand this behavior. From \eqref{eq:SPhi}, we can find the
tunneling factor by evaluating
\begin{equation}
S\left(\epsilon,\rho_{0}\right)=\int_{\epsilon}^{\rho_{0}}d\rho\sqrt{\fft{\nu_{2}^{2}}{\rho^{2}}+\fft{\vec{k}^{2}L/2}\rho-\omega^{2}}.
\end{equation}
In the near boundary region $\epsilon\ll\rho_{*}$, the integral will
just generate the expected power-law behavior $\epsilon^{2\nu_{2}}$,
which is stripped off by the factor $\epsilon^{-2\nu_{2}}$ in \eqref{eq:chiWKB}.
For large $\alpha$, the tunneling region will contribute an additional
term of order
\begin{equation}
S\sim k\int_{\rho_{*}}^{\rho_{0}}d\rho\sqrt{\fft{L/2}\rho}\approx k\int_{0}^{\frac{\vec{k}^{2}L}{2\omega^{2}}}d\rho\sqrt{\fft{L/2}\rho}\sim\frac{\vec{k}^{2}L}{\omega}\sim\alpha,
\end{equation}
and the actual suppression term is $\sim e^{-\alpha}$ instead of
just $e^{-|\vec{k}|}.$ This result has a simple interpretation: For
a finite tunneling region $[R_{1},R_{2}]$ , the barrier can be made
arbitrarily high by taking $|\vec{k}|\rightarrow\infty$, resulting
in exponential suppression $e^{-|\vec{k}|}$. However, for pure Lifshitz,
the tunneling region can also be made arbitrarily wide by taking either
$\omega\rightarrow0$ at fixed $|\vec{k}|$, or $|\vec{k}|\rightarrow\infty$
at fixed $\omega$. Since the WKB functional $S$ is a measure for
the area between the wavefunction and the tunneling
potential, we end up with a suppression in $\alpha\sim|\vec{k}|\cdot({|\vec{k}|}/{\omega})$.

\begin{figure}
\begin{centering}
\hspace{-0.5cm}\includegraphics[width=7.5cm]{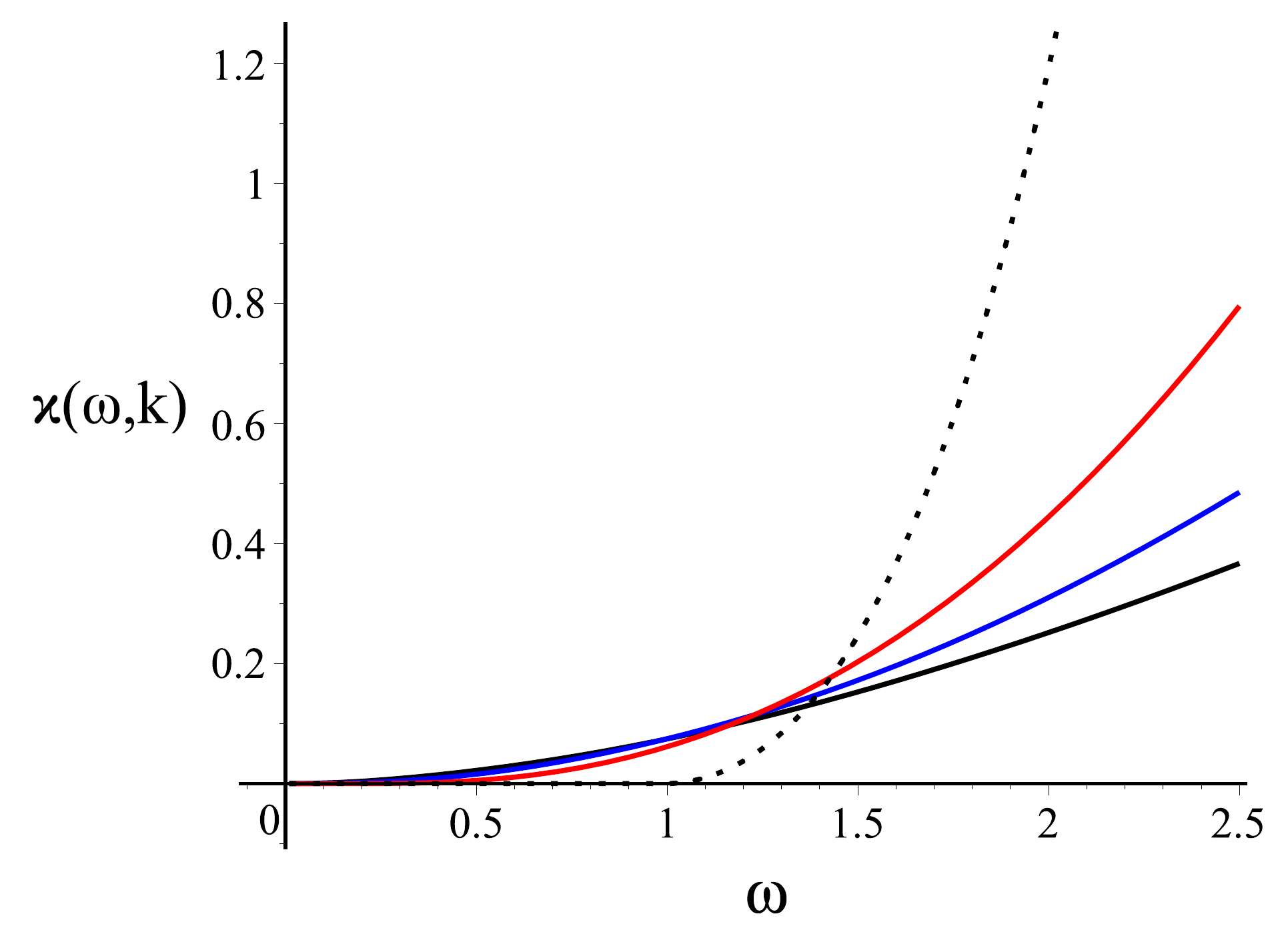}~~~\includegraphics[width=7.5cm]{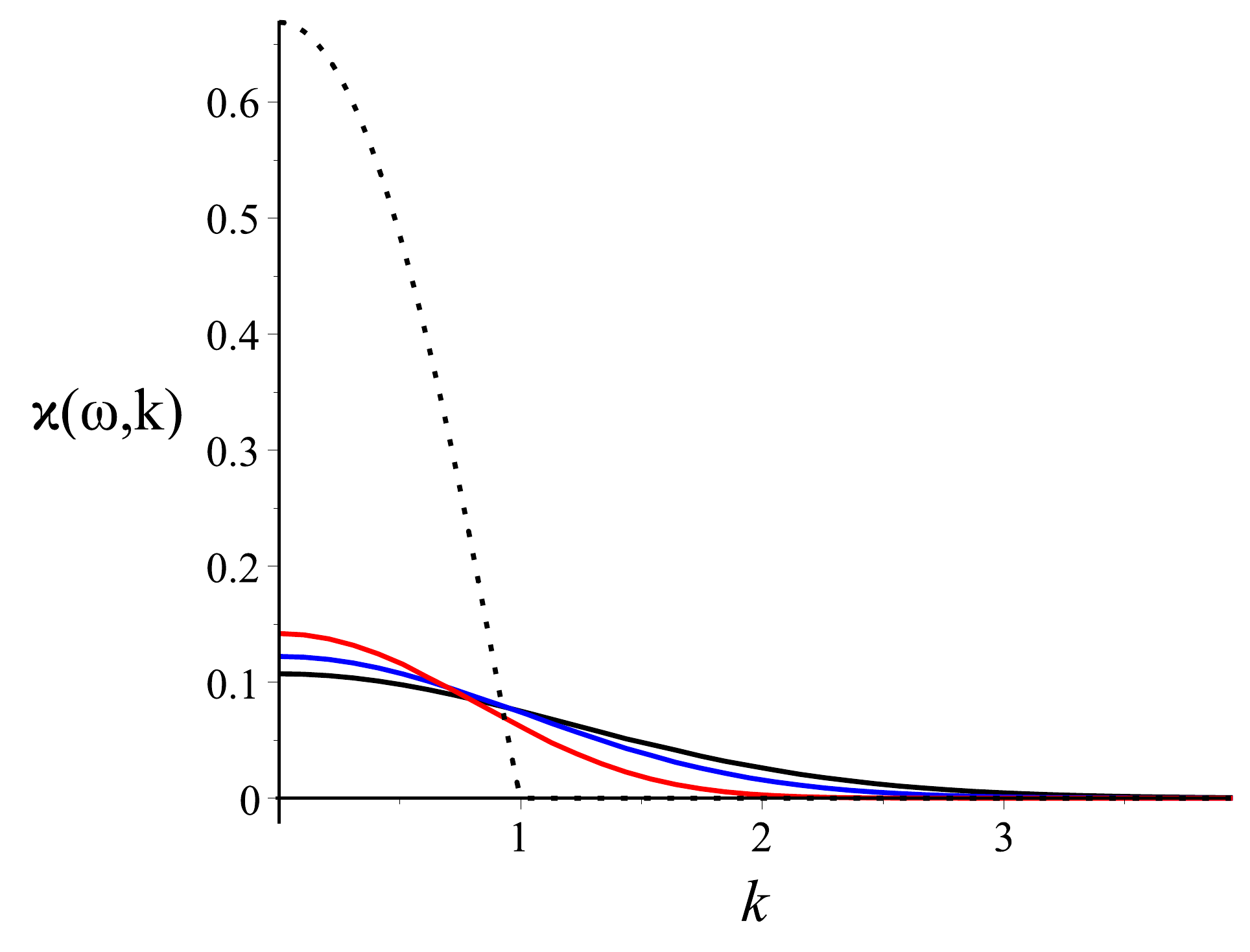}
\par\end{centering}
\caption{\label{fig:speclif}Plot of the spectral function $\chi(\omega,\vec{k})$
for Lifshitz with $z=2,3,4$ (red, blue, black). The AdS spectral
function is shown as a dotted line. Left: Varying $\omega$ while keeping
$|\vec{k|}=1/L$ fixed. Right: Varying $|\vec{k}|$ while keeping $\omega=1/L$
fixed. }
\end{figure}

To demonstrate that similar results hold for Lifshitz with general
$z$, we also computed the spectral function $\chi$ for $z=2,3,4$
numerically. Figure \ref{fig:speclif} shows plots of the spectral
function as a function of $\omega$ and $|\vec{k}|$ respectively.
For AdS, modes with spacelike momenta $|\vec{k}|^{2}>\omega^{2}$
have zero spectral weight. For Lifshitz, however, we can clearly see
an exponential tail both at small $\omega$ and large $|\vec{k}|$
due to tunneling, indicating the by now familiar insensitivity to
horizon boundary conditions. From the WKB approximation (\ref{eq:chiWKB}), we
expect the asymptotic behavior $\chi\sim\mathrm{exp}\left(-\lambda\alpha^{{1}/{\zeta}}\right)$,
with
\begin{equation}
\lambda=\fft{\sqrt{\pi}\Gamma(1/\zeta-1/2)}{2\Gamma(1/\zeta)},\qquad\alpha=\left(\fft{\omega L}z\right)^{\zeta}\biggl(\fft{\vec{k}}\omega\biggr)^{2},\qquad \zeta=2\left(1-\frac{1}{z}\right).
\end{equation}
Our numerical results confirm this behavior (see Table~\ref{tab:Knum}
for best-fit values).

\begin{table}
\begin{centering}
\begin{tabular}{|c|c|c|}
\hline
$z$  & $2/\zeta$  & $2/\zeta_{\mathrm{num}}$\tabularnewline
\hline
\hline
2  & 2  & 2.05\tabularnewline
\hline
3  & $3/2$  & 1.55\tabularnewline
\hline
4  & $4/3$  & 1.39\tabularnewline
\hline
\end{tabular}
\par\end{centering}

\caption{\label{tab:Knum}Best fit results for numerically obtained spectral
functions.}
\end{table}

\section{\label{sec:rgflows}Spectral functions for Lorentz-breaking RG flows}

Our discussion so far has been focused on the insensitivity of the
Green's function to a change of horizon boundary conditions. The goal
of this section is to reformulate this statement in a more physical
way. We do this by showing that for spacetimes with a tunneling barrier,
the retarded Green's function is in fact exponentially insensitive
to the near-horizon geometry itself. In terms of the corresponding
RG flow, this has the somewhat surprising consequence that in the
low energy, large momentum limit, the spectral function shows a universal
behavior that depends only very weakly on the details of the IR theory.
In that sense, flows with different IR fixed points are almost non-distinguishable.

To see explicitly how this arises, consider the case of an RG flow
that interpolates between two different fixed points in the UV and IR. Since
the dual spacetime interpolates between two different geometries at the
horizon and the boundary, we introduce $\rho_c$ as a cross-over scale between
these two asymptotic geometries, and split the effective potential as
\begin{equation}
U=\begin{cases}
U_{\mathrm{UV}}, & \rho\ll\rho_{c}\\
U_{\mathrm{IR}}, & \rho\gg\rho_{c}.
\end{cases}
\end{equation}
Although the potential near $\rho_{c}$ depends on the precise
way these two geometries are glued together, we will not need
to know its explicit form in the intermediate region in order to study
the general behavior of the spectral function. To simplify our discussion,
let us assume that $U$ decreases monotonically, so that there are no bound states, and
that $U_{\mathrm{UV}}\left(\rho\rightarrow0\right)\sim({\nu^{2}-1/4})/{\rho^{2}}$,
as before.

We would like to extract information about IR physics from the spectral
function. First, consider frequencies $\omega$ large enough so that the classical turning
point $\rho_0(\omega)$ is in the UV, $\rho_{0}\left(\omega\right)\ll\rho_{c}$. Physically, since we are probing the geometry at high energies, $\chi$ is completely
independent of the IR geometry.  All that remains is the spectral function for the dual theory at the UV fixed point.

 Next, let us use the WKB approximation to see what happens when
we lower the energy far enough that the scalar wavefunction actually
has to tunnel through part of the IR-potential, {\it i.e.}\ $\rho_{0}\left(\omega\right)\gg\rho_{c}$.
We can approximate the WKB-integral as
\begin{equation}
S\left(\rho,\rho_{0}\right)\approx S_{\mathrm{UV}}\left(\rho,\rho_{c}\right)+S_{\mathrm{IR}}\left(\rho_{c},\rho_{0}\right)+\cdots,
\end{equation}
with $S_{\mathrm{UV/IR}}=\int d\rho\sqrt{U_{\mathrm{UV/IR}}-\omega^{2}}$.
Here the ellipsis denotes terms that depend on the precise way the
two geometries are glued together. The spectral function now becomes
\begin{equation}
\chi\approx K\epsilon^{-2\nu}e^{-2S_{\mathrm{UV}}\left(\epsilon,\rho_{c}\right)}e^{-2S_{\mathrm{IR}}\left(\rho_{c},\rho_{0}\right)},\label{eq:chiUVIR}
\end{equation}
and the information about IR physics shows up in the factor $e^{-2S_{\mathrm{IR}}}$.
For relativistic flows, one roughly gets $\chi\sim f\left(\omega\right){\cal O}\left(1\right)e^{-2S_{\mathrm{IR}}\left(\rho_{c},\rho_{0}\right)}$,
and the IR geometry has an ${\cal O}(1)$ imprint on the spectral
function. However, as we saw previously, if the UV fixed point has a
Lifshitz scaling symmetry, the tunneling barrier will induce an exponential
factor and we get
\begin{equation}
\chi\sim f\left(\omega\right){\cal O}\left(e^{-c|\vec{k}|}\right)e^{-2S_{\mathrm{IR}}\left(\rho_{c},\rho_{0}\right)}.
\end{equation}
At large $\vec{|k}|$, all the information about IR physics is hidden
under an exponentially small factor. In the limit $|\vec{k}|\rightarrow\infty$,
a change of the geometry in the deep IR has no effect on the spectral
function.

The factorization of $\chi$ into UV and IR factors in \eqref{eq:chiUVIR}
allows us to make an even more general statement: Consider any flow
that breaks $(d+1)$-dimensional Lorentz-invariance somewhere in the
bulk, {\it i.e.}\ $A\neq B$ in \eqref{eq:effpot}. At low frequencies $\omega$
and large momenta $|\vec{k}|$ the spectral function will have a universal
exponential damping factor $e^{-c|\vec{k}|}$ due to the tunneling
barrier $\vec{k}^{2}e^{2A-2B}$.

We can demonstrate this behavior explicitly by considering holographic
RG flows with AdS$_{2}\times\mathcal{\mathbb{R}}^{d}$ near-horizon
geometry. One important example of such spacetimes are extremal charged black branes in $\mathrm{AdS}_{d+2}$,
which are holographically dual to theories at finite charge density \cite{Chamblin:1999tk}. Placing fermions on this background
allows us to study Fermi surfaces in non-Fermi liquids \cite{Lee:2008xf,Basu:2009qz,Faulkner:2009wj,Faulkner:2011tm}.
AdS$_{2}\times\mathcal{\mathbb{R}}^{d}$ also plays a crucial role
in the resolution of the tidal singularity in Lifshitz spacetime \cite{Harrison:2012vy,Bhattacharya:2012zu,Knodel:2013fua,Barisch-Dick:2013xga}.

Of particular interest to us are flows with either $\mathrm{AdS}_{d+2}$
or $\mathrm{Lif}_{z}$ near-boundary behavior. For both cases, the
Schr\"odinger potential can be written as
\begin{equation}
U=\begin{cases}
\frac{\nu_{z}^{2}-{1}/{4}}{\rho^{2}}+\vec{k}^{2}\left(\frac{L}{z\rho}\right)^{2-2/z}, & \rho\ll\rho_{c};\\
\frac{\nu_{\mathrm{\infty}}^{2}-{1}/{4}}{\rho^{2}} ,& \rho\gg\rho_{c},
\end{cases}
\end{equation}
where $\nu_{\infty}^{2}=\left(mL_{\mathrm{IR}}\right)^{2}+\vec{k}^{2}L_{\mathrm{IR}}+{1}/{4}$,
and the null energy condition requires $z\geq1$. The near-horizon
AdS$_{2}\times\mathbb{R}^{d}$ itself has a holographic dual, which
is a $\mathrm{CFT}_{1}$. In particular, there is a corresponding
spectral function
\begin{equation}
\chi_{\mathrm{cft}}\approx K\epsilon^{-2\nu_{\infty}}e^{-2S_{\mathrm{IR}}\left(\epsilon,\rho_{0}\right)}.
\end{equation}
Again, in the high energy limit the spectral function carries no information
about the IR CFT. At low energies, specifically $\omega\ll{\nu_{\infty}}/{\rho_{c}}$
or equivalently $\rho_{c}\ll\rho_{0}\left(\omega\right)$, we can
derive a direct relation between $\chi_{\mathrm{cft}}$ and the full
spectral function:
\begin{equation}
\chi\left(\omega\ll\frac{\nu_{\infty}}{\rho_{c}},\vec{k}\right)\approx\epsilon^{-2\nu}e^{-2S_{\mathrm{UV}}\left(\epsilon,\rho_{c}\right)}\chi_{\mathrm{cft}}.\label{eq:chivschift}
\end{equation}
Let us evaluate this expression for large $|\vec{k}|.$ The integral
we have to perform is
\begin{equation}
S_{\mathrm{UV}}\left(\rho,\rho_{c}\right)=\int_{\rho}^{\rho_{c}}\sqrt{\frac{\nu^{2}}{\rho^{2}}+\vec{k}^{2}\left(\frac{L}{z\rho}\right)^{2-2/z}-\omega^{2}}.\label{eq:SUV}
\end{equation}
The crossover scale from ${1}/{\rho^{2}}$ behavior to ${1}/{\rho^{2(1-1/z)}}$ behavior
is at $\rho_{*}\equiv\left({\nu}/{p}\right)^{z}$. We will assume
that ${|\vec{k}|\rho_{c}^{{1}/{z}}}/{\nu}\gg1$, so that this
crossover still happens in the UV region, i.e. $\rho_{*}\ll\rho_{c}.$
Since $\omega\ll{\nu_{\mathrm{\infty}}}/{\rho_{c}}$, and the
momentum is taken to be large, we can simply neglect the $\omega^{2}$ term
in \eqref{eq:SUV}. Introducing the new variable $u\equiv({p^{2}}/{\nu^{2}z^{2(1-1/z)}})\rho^{2/z}$,
we can evaluate the integral:
\begin{equation}
S_{\mathrm{UV}}\left(\rho,\rho_{c}\right)\approx\frac{z\nu}{2}\left[2\sqrt{1+u}+\log\frac{\sqrt{1+u}-1}{\sqrt{1+u}+1}\right]_{u}^{u_{c}}.
\end{equation}
Expanding this result around large $u_{c}$ and small $u$, we find
\begin{equation}
e^{-2S_{\mathrm{UV}}\left(\epsilon,\rho_{c}\right)}\approx\epsilon^{2\nu}\left(\frac{|\vec{k}|}{2\nu z^{1-1/z}}\right)^{2z\nu}e^{-2\left(z\rho_{c}\right)^{\frac{1}{z}}|\vec{k}|}.
\end{equation}
Plugging this back into \eqref{eq:chivschift}, we see that the $\epsilon$-dependent
terms precisely cancel, and we are left with
\begin{equation}
\chi\big(\omega\ll\frac{\nu_{\infty}}{\rho_{c}},|\vec{k}|\gg\frac{\nu}{\rho_{c}^{1/z}}\big)\approx
K\left(\frac{|\vec{k}|}{2\nu z^{1-1/z}}\right)^{2z\nu}e^{-2\left(z\rho_{c}\right)^{\frac{1}{z}}|\vec{k}|}\chi_{\mathrm{cft}}.\label{eq:specUVIR}
\end{equation}
The spectral function at low energies is directly proportional to
the IR spectral function $\chi_{\mathrm{cft}}$. At large $|\vec{k}|$,
$\chi$ is exponentially small. It might seem surprising that this
is true even for the case of asymptotically AdS spacetimes, where
$z=1$. As was discussed in \cite{Keeler:2013msa}, this is because
even though pure AdS does not have a tunneling barrier, flowing to
a non-relativistic $\mathrm{AdS_{2}}\times\mathbb{R}^{d}$ horizon
necessarily breaks Lorentz invariance and introduces a tunneling barrier.

The relation \eqref{eq:specUVIR} between UV and IR spectral functions has been obtained previously, using standard matching techniques \cite{Faulkner:2009wj}.
Our calculation sheds new light on this result: While the spectral function is dominated by IR physics at low $\omega$, the numerical coefficient relating $\chi$ and $\chi_{\mathrm{cft}}$ is exponentially small at large momenta. For a boundary observer, the signature of low-energy physics is hidden under an exponential tail. 

\section{Comments on Schr\"odinger spacetimes}
\label{sec:schroedinger}

In addition to spacetimes exhibiting non-relativistic Lifshitz scaling symmetry, there has also been much interest in spacetimes that realize the non-relativistic conformal group, or Schr\"{o}dinger group \cite{Son:2008ye,Balasubramanian:2008dm,Adams:2008wt,Blau:2009gd,Blau:2010fh}.  The metric
of Schr\"odinger spacetime is
\beq\label{eq:Schrmetric}
ds^2_{d+3} = -\frac{dt^2}{r^{2z}}+ \frac{2 d\xi\, dt + d\vec{x_d}^2+dr^2}{r^2},
\eeq
where $r=0$ is the UV boundary, and $z$ is again the dynamical exponent; we have additionally set $L=1$ for simplicity.  (Strictly speaking,
the Schr\"odinger group with special conformal generator is only realized for $z=2$.)
Here $\xi$ represents an additional null direction; as momentum $P_\xi\equiv M$ along this auxiliary direction is related to particle number, we work in a fixed superselection sector for $M$.

Although this spacetime is not of the general form (\ref{eq:lifgmet}) that we studied previously, analysis of the scalar behavior proceeds quite similarly.  Making the ansatz
\beq
\phi(t,\xi,\vec{x},r) = r^{\frac{d+1}{2}}e^{i(\omega t-M\xi+\vec{k}\cdot\vec{x})}\psi(r),
\eeq
we can rewrite the Klein-Gordon equation $(\square-m^2)\phi=0$ as the effective Schr\"odinger-like equation
\beq\label{eq:schr-schrlike}
-\psi''(r)+U(r)\psi(r)=2M \omega \psi(r).
\eeq
Here the effective potential is
\beq\label{eq:Uschr}
U= \frac{(d+2)^2/4+m^2-1/4}{r^2}+k^2+M^2 r^{2-2z}.
\eeq
As in the Lifshitz case, we have chosen our radial coordinate in (\ref{eq:Schrmetric}) so that there is a clear effective energy term in the effective Schr\"odinger equation (\ref{eq:schr-schrlike}).  In the Schr\"odinger spacetime however, this effective energy term is $2M \omega$.

Again as in the Lifshitz case, setting $z=1$ in the metric (\ref{eq:Schrmetric}) simply reproduces AdS$_{d+3}$ spacetime in a light-like coordinate system.  Other values of $z$ will produce different behavior from the Lifshitz case; we will consider both $z=2$ and $1<z<2$.

\subsection{Schr\"odinger spacetime with $z=2$}\label{sec:schrz2}
We begin with $z=2$ Schr\"odinger space.  In this case, the effective potential becomes
\beq\label{eq:Uschr,z=2}
U = \frac{\nu^2-1/4}{r^2}+k^2,
\eeq
where
\beq
\nu^2=(d+2)^2/4+m^2 +M^2.
\label{eq:nus2}
\eeq
The effective potential here takes the same form as in AdS$_{d+3}$, except the mass has been shifted by $m^2 \rightarrow m^2+M^2$.  The potential contains only a constant term and a $1/r^2$ term.  There is thus no tunneling regime; the $1/r^2$ potential merely provides the polynomial scaling $\phi\sim r^{\Delta}$ near the $r\rightarrow 0$ boundary.

Accordingly, in contrast with $z=2$ Lifshitz, the smearing function here can be defined; the computation is similar to the AdS case, except for the mass shift and the effective energy change. 
The scalar Green's function is also easy to compute; we obtain the AdS result given in (\ref{eq:AdSG}), with the replacement $q\rightarrow \sqrt{2M\omega -\vec{k}^2}$, and with $\nu$
given in (\ref{eq:nus2}) \cite{Son:2008ye,Balasubramanian:2008dm}:
\begin{equation}
G_{R}(\omega,\vec{k},M)=K\frac{\Gamma(-\nu)}{\Gamma(\nu)}
\left(\frac{2M\omega-\vec k^2}{4}\right)^{\nu}e^{-i\nu\pi}.
\label{eq:z=2schG}
\end{equation}
Correspondingly the dispersion relation is $\omega =\vec{k}^2/2M$, which is manifestly non-relativistic.
For $\omega <\vec{k}^2/2M$, the spectral function is exactly zero.
Note that it is exactly at $z=2$ when the Schr\"odinger algebra gains a special conformal generator.
Moreover, there is no tunneling regime, and correspondingly the spectral function has no
exponentially suppressed region.

\subsection{Schr\"odinger spacetime with $1<z<2$}

For $1<z<2$, Schr\"odinger spacetime has an effective potential of the same functional form as that for Lifshitz, (\ref{eq:LifU}), but with the identification $z_L=1/(2-z)$.  In this case, a tunneling barrier will
be present and will affect the low frequency modes.  The mapping to Lifshitz allows us to read
off the exact solution for the $z=3/2$ Schr\"odinger Green's function from that for Lifshitz with
$z_L=2$ in (\ref{eq:GLif2}):
\beq\label{eq:Gschz3over2}
G_{S}(\omega,\vec{k},M)=K(2q)^{2\nu_s}e^{-i\nu\pi}\frac{\Gamma(-2\nu)}{\Gamma(2\nu)}
\frac{\Gamma\left(\frac{1}{2}+\nu+i\frac{M^2}{2q}\right)}{\Gamma\left(\frac{1}{2}-\nu+i\frac{M^2}{2q}
\right)},
\eeq
where we have defined
\beq
\nu = \sqrt{(d+2)^2/4+m^2}\quad\mbox{and}\quad
q =\sqrt{2M\omega-\vec{k}^2}.
\eeq
There are a few important differences in the physics.  First, in the Schr\"odinger case we are usually interested in a fixed superselection sector for $M$, as it represents particle number.  As such, to produce a smearing function in position space, we would not integrate over the $M$ momentum. Since it is $M$ which controls the size of the tunneling region, the smearing function will be mathematically definable.  Of course since we work with modes of fixed $M\equiv P_\xi$ momentum they are intrinsically not local in the $\xi$ direction, so from boundary data we cannot reconstruct $\xi$ locality anyhow. 

Next, the potential in Schr\"odinger space, (\ref{eq:Uschr}), reaches a minimum set by the spatial momentum $|\vec{k}|^2$, rather than the $0$ reached by $U_{\text{Lif}}$.  Thus in order to have an allowed mode in the IR region, we must have effective energy satisfying $2M\omega >k^2$.  Consequently, for smaller $\omega$, the Green's function will have zero imaginary part.  
Lastly, within the allowed region, this spectral function will be exponentially suppressed in $M^2/2|q|$.
Thus, for $\vec{k^2}$ close to $2M\omega$ such that $|q|\ll M^2$, the spectral function is small, and the Green's function correspondingly becomes insensitive to changes in IR boundary conditions.

\section{Analytic properties of the Green's function}
\label{sec:analytic}\
In order to put our results into context, it is worth recalling that field theory Green's functions
exhibit a rich structure in the complex $\omega$-plane.  At zero temperature, one typically
defines three functions, namely the retarded, advanced, and time-ordered (causal) Green's functions
\begin{eqnarray}
G_R(\vec x,t;\vec x',t')&=&i\langle[\phi(\vec x,t)\phi(\vec x',t')]\rangle\Theta(t-t'),\nn\\
G_A(\vec x,t;\vec x',t')&=&-i\langle[\phi(\vec x,t)\phi(\vec x',t')]\rangle\Theta(t'-t),\nn\\
G_c(\vec x,t;\vec x',t')&=&i\langle T\phi(\vec x,t)\phi(\vec x',t')\rangle.
\end{eqnarray}
When Fourier transformed into $(\omega,\vec k)$, unitarity and causality imply that $G_R$
is analytic in the upper half of the complex $\omega$-plane, while $G_A$ is analytic in the
lower half.  These functions are not independent, but may be related by
\begin{equation}
G_R(\omega,\vec k)=[G_A(\omega,\vec k)]^*,
\end{equation}
as well as
\begin{equation}
G_c(\omega,\vec k)=G_R(\omega,\vec k)\theta(\omega)+G_A(\omega,\vec k)\theta(-\omega).
\end{equation}
In general, these Green's functions can be obtained from a single real analytic function
$G(\omega,\vec k)$ satisfying $[G(\omega,\vec k)]^*=G(\omega^*,\vec k)$ (except for possible
poles and branch cuts) by using an $i\epsilon$ prescription
\begin{eqnarray}
G_R(\omega,\vec k)&=&G(\omega+i\epsilon,\vec k),\nn\\
G_A(\omega,\vec k)&=&G(\omega-i\epsilon,\vec k),\nn\\
G_c(\omega,\vec k)&=&G(\omega+i\epsilon\mathrm{\,sign\,}\omega,\vec k).
\end{eqnarray}
The substitution for the time-ordered Green's function is equivalent to taking
$\omega^2\to\omega^2+i\epsilon$.  For real $\omega$, the spectral function is then given by
\begin{equation}
\chi(\omega,\vec k)=2\,\mathrm{Im}\,G_R(\omega,\vec k)=-i[G_R(\omega,\vec k)-G_A(\omega,\vec k)]
=-i[G(\omega+i\epsilon,\vec k)-G(\omega-i\epsilon,\vec k)].
\end{equation}
What this demonstrates is that non-vanishing spectral weight $\chi(\omega,\vec k)$ is related to
either poles or discontinuities across any branch cuts that lie on the real $\omega$ axis.  

These features are of course well known in field theory, so it is interesting to see how they
arise in the holographic Green's function computation.  For a bulk scalar in AdS or Lifshitz,
the Klein-Gordon equation, and hence effective Schr\"odinger-like equation (\ref{eq:schr}), is
quadratic in $\omega$.  However, the $\omega\to-\omega$ symmetry is broken
by imposing infalling boundary conditions at the horizon.  In other words, the holographic
computation directly gives $G_R$ without the need for any $i\epsilon$ prescription.  Nevertheless,
it is possible to analytically continue the resulting expressions to obtain $G(\omega,\vec k)$
in the complex $\omega$-plane.

\begin{figure}[t]
\begin{minipage}[c]{.45\linewidth}
\begin{subfigure}[b]{\linewidth}
\centering
\includegraphics[width=\linewidth]{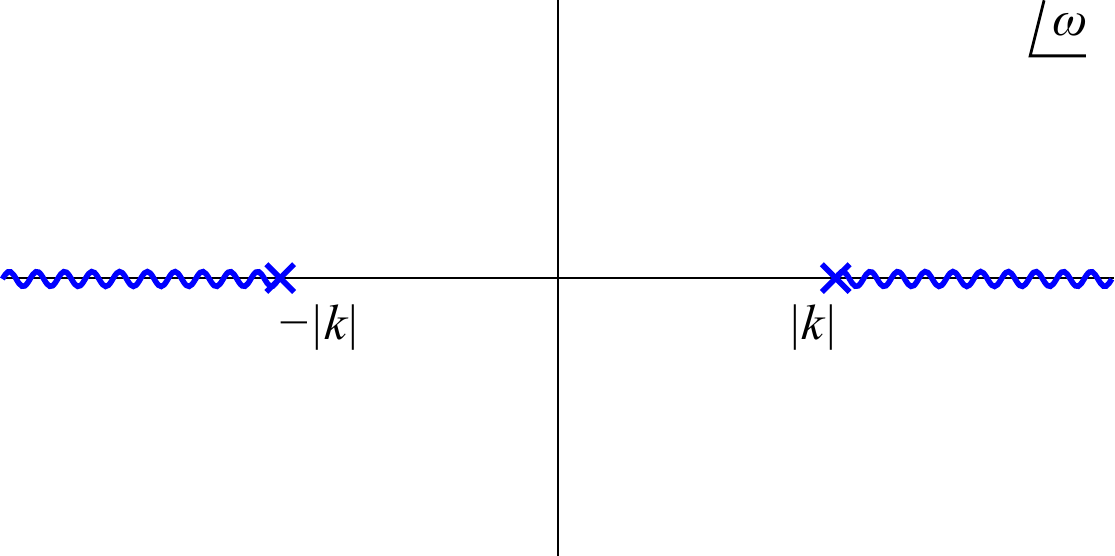}
\caption{Branch cut structure in the complex $\omega$-plane.\label{fig:z=1a}}
\end{subfigure}
\begin{subfigure}[b]{\linewidth}
\centering
\includegraphics[width=\linewidth]{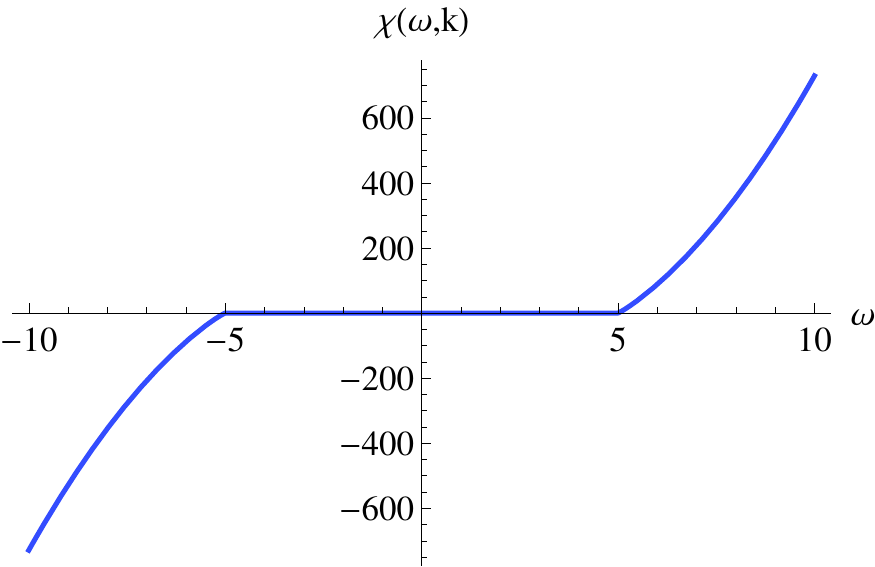}
\caption{Spectral function $\chi(\omega,|\vec k|)$ for $|\vec k|=5$.}
\end{subfigure}
\end{minipage}
\begin{subfigure}[c]{.545\linewidth}
\centering
\includegraphics[width=\linewidth]{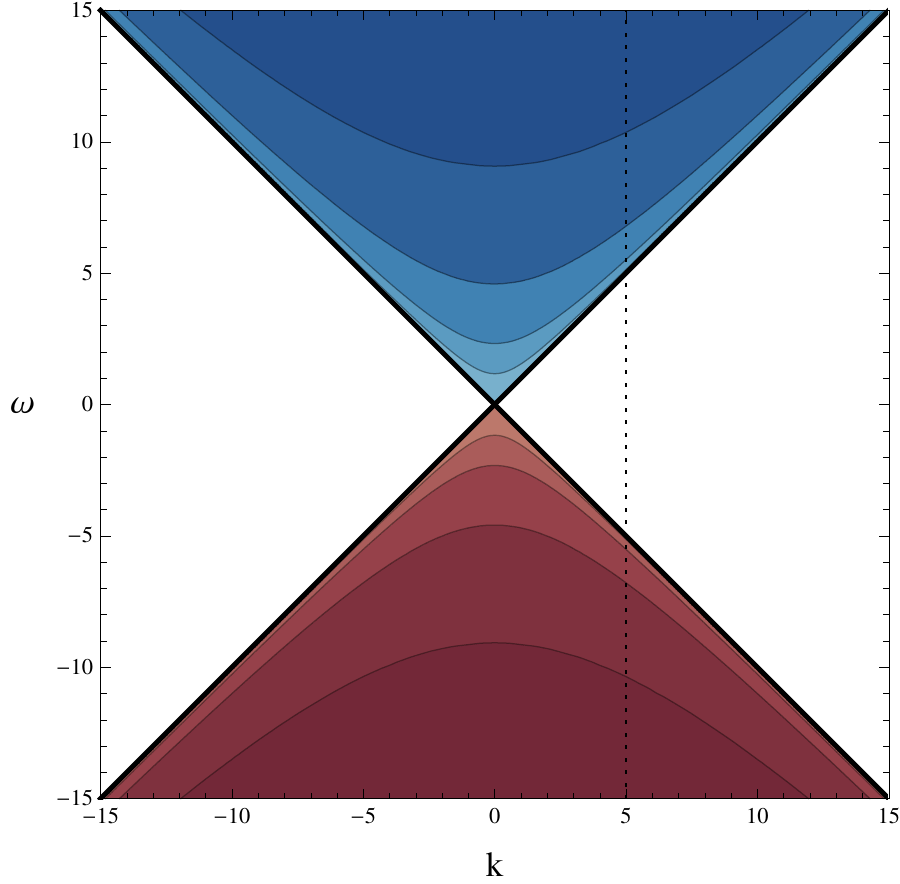}
\caption{Contour plot of $\chi(\omega,|\vec k|)$.  Note that the contour steps are logarithmic.  The dotted
line at $|\vec k|=5$ corresponds to the slice shown in ($b$).}
\end{subfigure}
\caption{\label{fig:z=1} The spectral function for AdS (see (\ref{eq:GAdS}), with $\nu=1.1$).  Note that $\chi(\omega,|\vec k|)$
vanishes identically for $|\omega|<|\vec k|$.}
\end{figure}

As an example, we may start with the retarded AdS Green's function given by (\ref{eq:AdSG})
with $b/a=0$, and obtain
\begin{equation}
G(\omega,\vec k)=K\fft{\Gamma(-\nu)}{\Gamma(\nu)}\left(\fft{\vec k^2-\omega^2}4\right)^\nu,\label{eq:GAdS}
\end{equation}
for non-integer values of $\nu$; we have set $L=1$ for notational ease throughout this section.  This function has branch points at $\omega=\pm|\vec k|$, and
as long as we take the principal branch of $z^\nu$, the branch cuts will extend out as shown in
Fig.~\ref{fig:z=1a}.  As a result, the spectral weight must vanish for $|\omega|<|\vec k|$.  This region
corresponds to the `energy' $\omega^2$ lying completely under the AdS effective potential given by
(\ref{eq:LifU}) with $z=1$.  In this case, the radial wavefunction never oscillates, and can be
chosen to be real, which is consistent with the vanishing of $\chi(\omega,\vec k)$.  Furthermore, in this
case there is no longer any freedom to modify the horizon boundary conditions, as one can only physically
choose the exponentially decaying solution at the horizon.  We give an example of the spectral function
for AdS in Fig.~\ref{fig:z=1}.

\begin{figure}[t]
\begin{minipage}[c]{.45\linewidth}
\begin{subfigure}[b]{\linewidth}
\centering
\includegraphics[width=\linewidth]{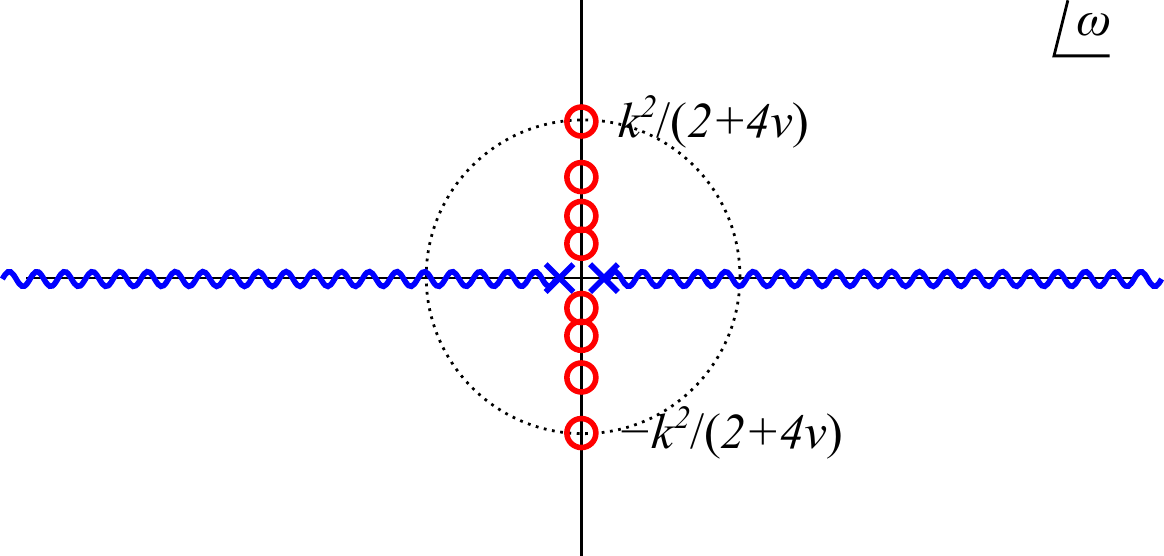}
\caption{Analytic features in the complex $\omega$-plane.}
\end{subfigure}
\begin{subfigure}[b]{\linewidth}
\centering
\includegraphics[width=\linewidth]{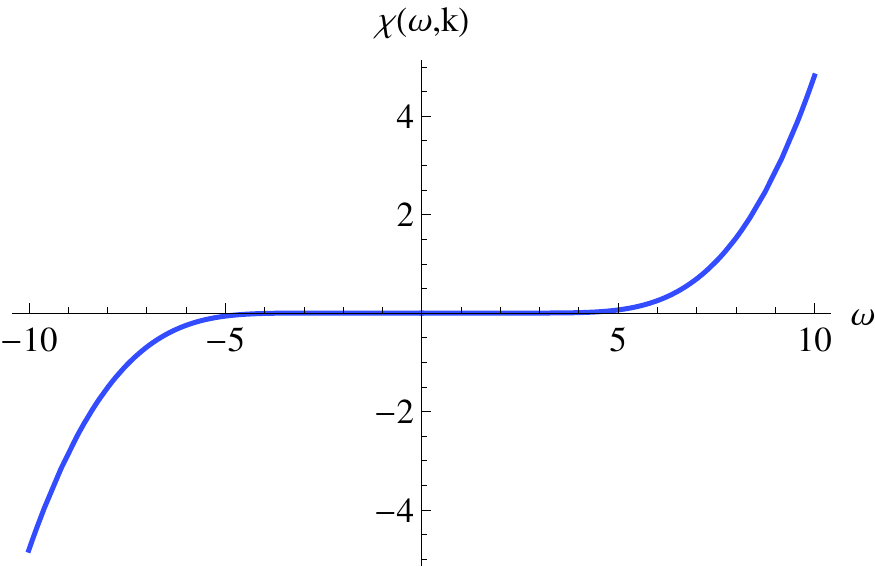}
\caption{Spectral function $\chi(\omega,|\vec k|)$ for $|\vec k|=5$.}
\end{subfigure}
\end{minipage}
\begin{subfigure}[c]{.545\linewidth}
\centering
\includegraphics[width=\linewidth]{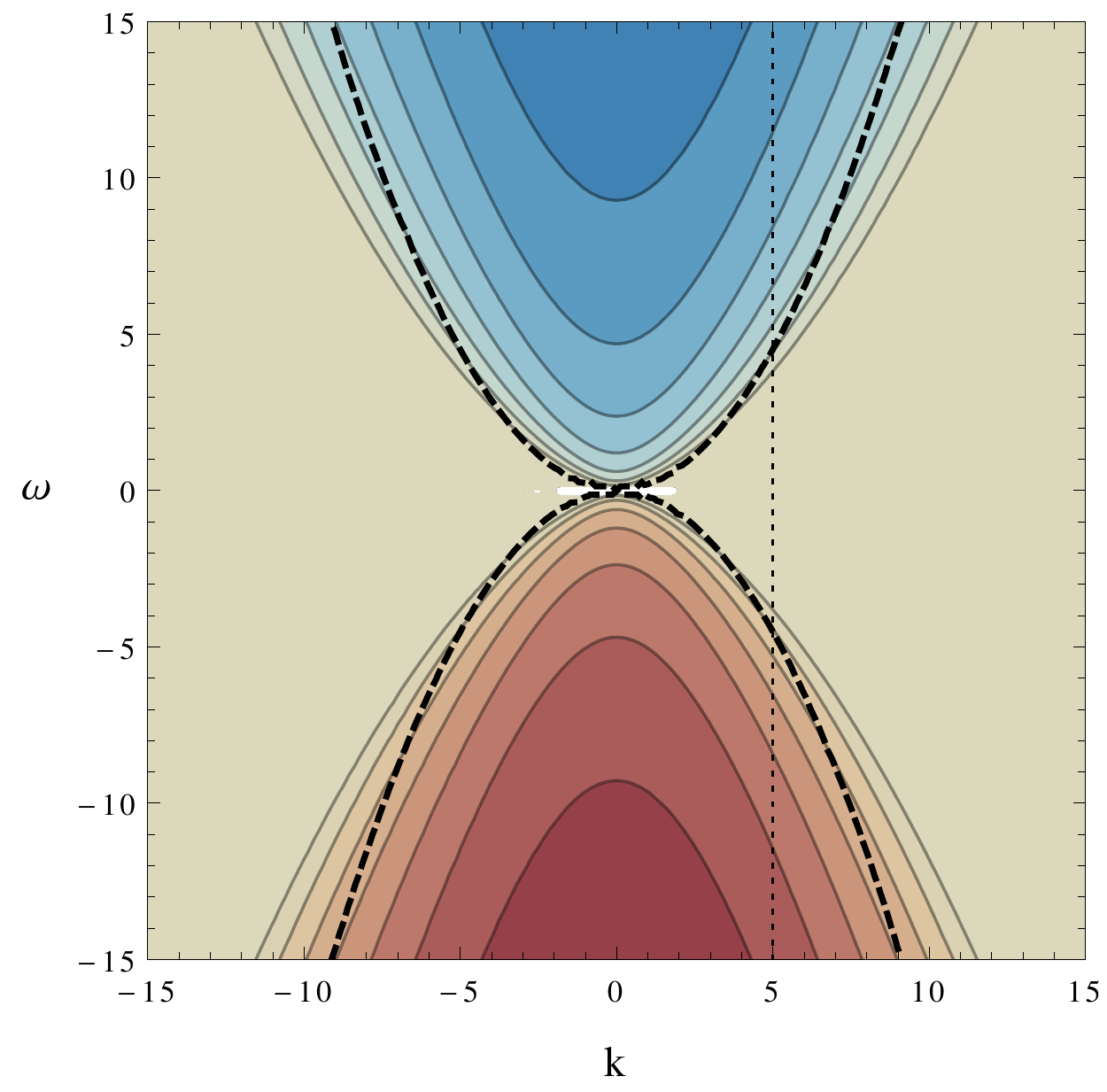}
\caption{Contour plot of $\chi(\omega,|\vec k|)$.  Note that the contour steps are logarithmic.  The dotted
line at $|\vec k|=5$ corresponds to the slice shown in ($b$). In the region outside of the dashed lines, $\omega \ll k^2/2\nu$, the spectral function is exponentially suppressed.}
\end{subfigure}
\caption{\label{fig:z=2} The spectral function for $z=2$ Lifshitz (see (\ref{eq:GLif2c}), with $\nu=1.1$).  In $(a)$, the branch cuts
extend from the origin to $\pm\infty$, and there are an infinite number of poles on the second sheet that
accumulate at the origin.  The spectral function is exponentially suppressed in the interior of the
dashed circle shown in $(a)$.}
\end{figure}

We now turn to the $z=2$ Lifshitz Green's function.  Starting from (\ref{eq:GLif2}),
we find the appropriate analytic continuation to be
\begin{equation}
G(\omega,\vec k)=K\fft{\Gamma(-2\nu_2)}{\Gamma(2\nu_2)}(-\omega^2)^{\nu_2}
\fft{\Gamma(\fft12+\nu_2+\vec k^2/4\sqrt{-\omega^2})}
{\Gamma(\fft12-\nu_2+\vec k^2/4\sqrt{-\omega^2})}.\label{eq:GLif2c}
\end{equation}
Again working with principal values, the factor $(-\omega^2)^{\nu_2}$ gives rise to a branch cut
running from the origin to $+\infty$ as well as from the origin to $-\infty$.  Thus $\chi$ is non-vanishing
for any $\omega\ne0$, although it becomes exponentially small for $|\omega|\ll\vec k^2/\nu_2$.
Note that, while the $\Gamma$-function in the numerator introduces poles in $G(\omega,\vec k)$,
they all lie on the unphysical second Riemann sheet. Even though they are in the second sheet, the accumulation of these poles causes an essential singularity at $\omega=0$. These features, along with the $z=2$
Lifshitz spectral function, are shown in Fig.~\ref{fig:z=2}.  In general, since the effective potential
$U(\rho)$ in (\ref{eq:LifU}) vanishes at the horizon for any $z>1$ Lifshitz geometry, the wavefunction
will be oscillatory at the horizon.  This in turn indicates that the retarded Green's function will be
complex, and hence that $\chi(\omega,\vec k)$ will be non-vanishing for any $\omega\ne0$.  Thus
the structure of branch cuts running along the positive and negative real $\omega$ axis is universal for
$z>1$ Lifshitz.

\begin{figure}[t]
\begin{minipage}[c]{.45\linewidth}
\begin{subfigure}[b]{\linewidth}
\centering
\includegraphics[width=\linewidth]{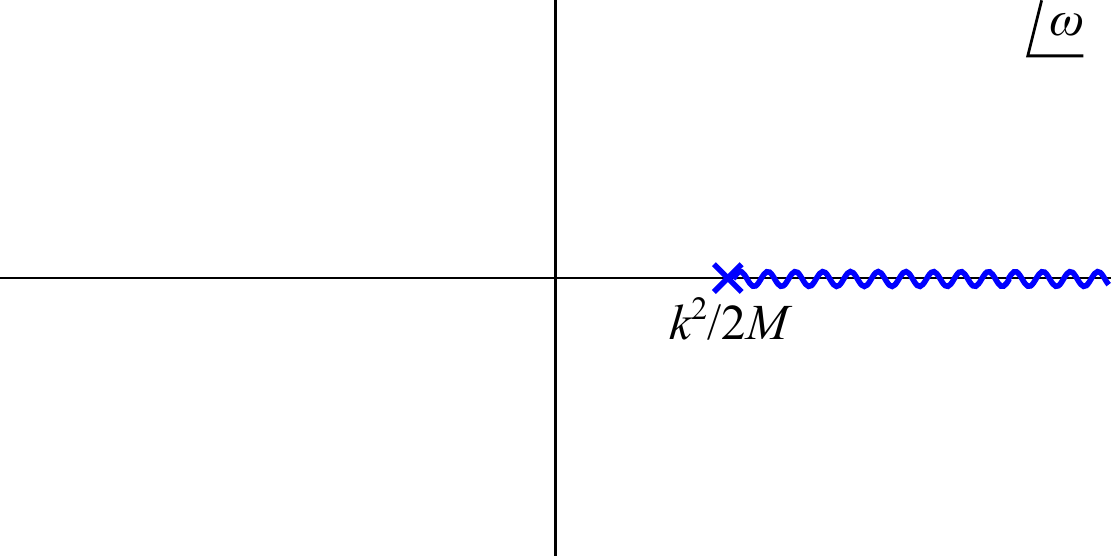}
\caption{Analytic features in the complex $\omega$-plane.}
\end{subfigure}
\begin{subfigure}[b]{\linewidth}
\centering
\includegraphics[width=\linewidth]{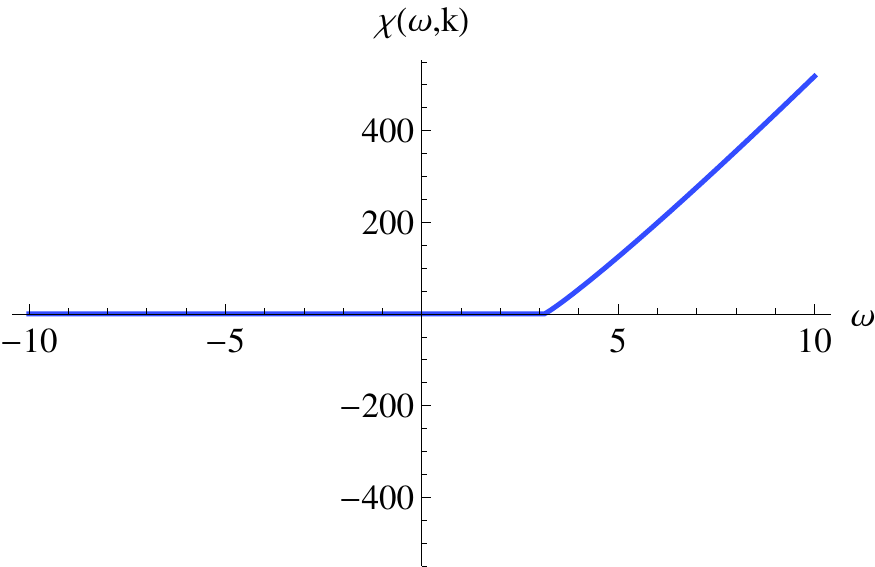}
\caption{Spectral function $\chi(\omega,|\vec k|)$ for $|\vec k|=5$.}
\end{subfigure}
\end{minipage}
\begin{subfigure}[c]{.545\linewidth}
\centering
\includegraphics[width=\linewidth]{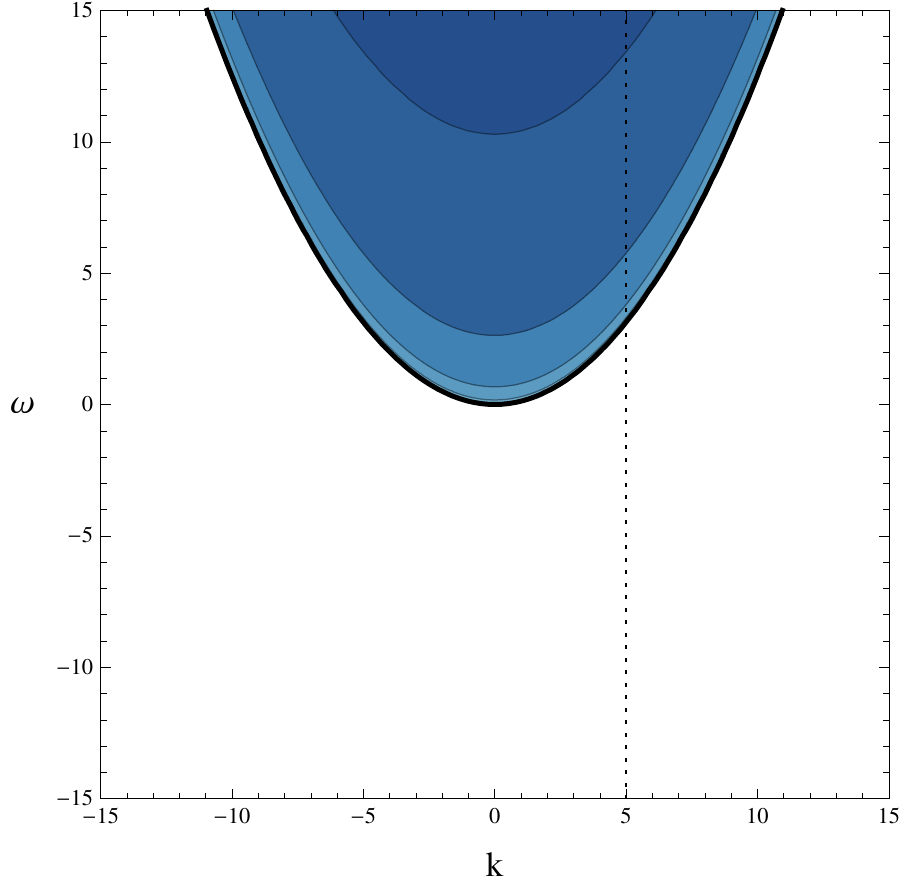}
\caption{Contour plot of $\chi(\omega,|\vec k|)$.  Note that the contour steps are logarithmic.  The dotted
line at $|\vec k|=5$ corresponds to the slice shown in ($b$). }
\end{subfigure}
\caption{\label{fig:z=2sch} The spectral function for $z=2$ Schr\"odinger (see (\ref{eq:GSchr2}) with $\nu=1.1$ and $M=4$). Note that $\chi$ vanishes exactly for $\omega < k^2/{2M}$.}
\end{figure}

In contrast with the Lifshitz backgrounds, the Schr\"odinger geometry breaks time reversal symmetry,
and as a result the Green's function will depend on $\omega$, and not its square.  For $z=2$
Schr\"odinger, the addition of the special conformal generator highly constrains the form of the
retarded Green's function to be that given in (\ref{eq:z=2schG}).  As in the AdS case, it is straightforward
to extend this into the complex $\omega$-plane.  We find
\begin{equation}
G(\omega,\vec k)=K\fft{\Gamma(-\nu)}{\Gamma(\nu)}\left(\fft{\vec k^2-2M\omega}4\right)^\nu.\label{eq:GSchr2}
\end{equation}
In this case, there is a single branch point at $\omega=\vec k^2/2M$, with a branch cut running
to $+\infty$.  This corresponds to the standard $z=2$ dispersion relation.  As discussed in section \ref{sec:schrz2}, the Schr\"odinger $z=2$ Green's function does not have a suppressed region, as there is no tunneling barrier in the effective potential.  Both of these features are shown in Figure \ref{fig:z=2sch}.

\begin{figure}[t]
\begin{minipage}[c]{.45\linewidth}
\begin{subfigure}[b]{\linewidth}
\centering
\includegraphics[width=\linewidth]{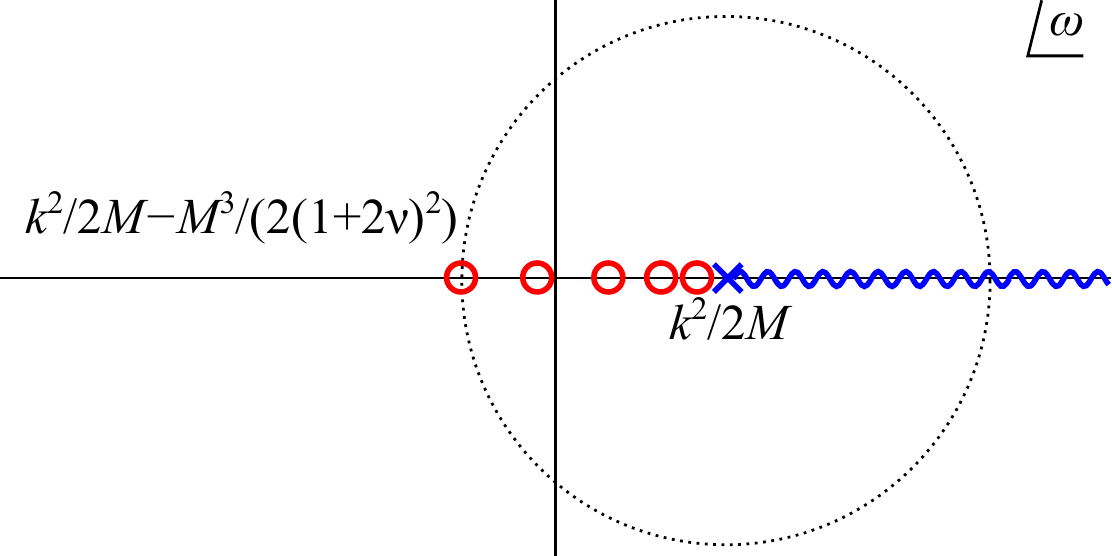}
\caption{Analytic features in the complex $\omega$-plane.}
\end{subfigure}
\begin{subfigure}[b]{\linewidth}
\centering
\includegraphics[width=\linewidth]{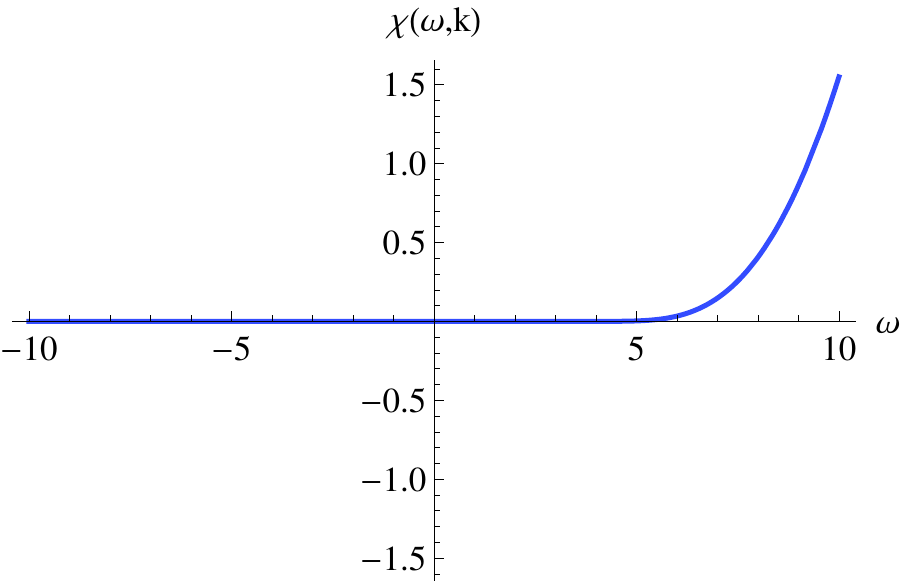}
\caption{Spectral function $\chi(\omega,|\vec k|)$ for $|\vec k|=5$.}
\end{subfigure}
\end{minipage}
\begin{subfigure}[c]{.545\linewidth}
\centering
\includegraphics[width=\linewidth]{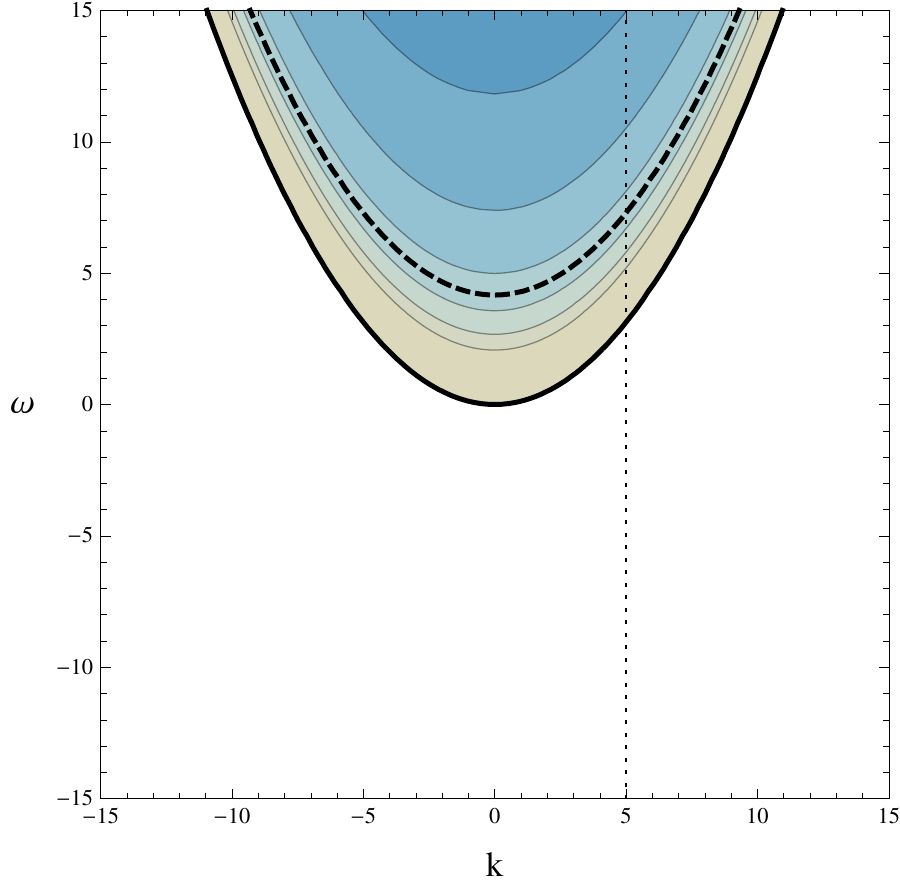}
\caption{Contour plot of $\chi(\omega,|\vec k|)$.  The contour steps are logarithmic.  The dotted
line at $|\vec k|=5$ corresponds to the slice shown in ($b$). Note that $\chi$ vanishes exactly for $\omega < k^2/{2M}$. In the region between the solid and dashed lines, $\omega-k^2/2M\ll M^3/\nu^2$, the spectral function is exponentially suppressed.}
\end{subfigure}
\caption{\label{fig:z=3over2} The spectral function for $z=3/2$ Schr\"odinger (see (\ref{eq:GSchr32}), with $\nu=1.1$ and $M=4$). 
In $(a)$, the branch cut extends from $\omega=k^2/2M$ to $+\infty$, and there are an infinite number
of poles on the second sheet that accumulate at the branch point. The spectral function is exponentially
suppressed in the interior of the dashed circle shown in $(a)$.}
\end{figure}

For $z=3/2$ Schr\"odinger, we find some similarities with both Lifshitz $z=2$ and Schr\"odinger $z=2$.  As in the Schr\"odinger $z=2$ case, time reversal symmetry is broken, and so the Green's function depends on $\omega$ (not $\omega^2$).  However, since there is no special conformal generator here, the form of the retarded Green's function is not heavily constrained; it can be derived from the Lifshitz $z=2$ result as shown in (\ref{eq:Gschz3over2}).  Again it can be extended into the complex plane, giving
\beq
G(\omega,\vec{k},M)=K(4(\vec k^2-2M\omega))^{\nu_s}\frac{\Gamma(-2\nu_s)}{\Gamma(2\nu_s)}
\frac{\Gamma\left(\frac{1}{2}+\nu_s+{M^2}/{2\sqrt{\vec{k}^2-2M\omega}}\right)}{\Gamma\left(\frac{1}{2}-\nu_s+{M^2}/{2\sqrt{\vec{k}^2-2M\omega}}\right)}.\label{eq:GSchr32}
\eeq
As in the Schr\"odinger $z=2$ case, the spectral function is exactly zero for $\omega<k^2/2M$.  There is a single branch point at $\omega<k^2/2M$ with a branch cut going to $+\infty$. However, as in Lifshitz, there is still a suppressed region; for $0<\omega-k^2/2M\ll M^3/\nu^2$, the spectral function is exponentially suppressed due to the tunneling potential.  Additionally, there are poles on the unphysical second sheet, and their accumulation causes the branch point at $\omega=k^2/2M$ to become an essential singularity. These features are depicted in Figure \ref{fig:z=3over2}.

\section{Discussion}\label{sec:discussion}
We have found a region of
momentum space ($\omega\ll1$, $|\vec{k|}\gg1$) where the holographic
Green's function of Lifshitz spacetime is exponentially insensitive
to a change of horizon boundary conditions. As we argued previously, this implies that the two-point function is insensitive to the geometry in the deep IR itself. 
Our discussion provides a new perspective on the problem of
finding the ``true'' IR endpoints of flows involving Lifshitz. It
has been shown that Lifshitz spacetime suffers from a tidal singularity
at the horizon, which if taken at face value leads to unphysical results
\cite{Copsey:2010ya,Horowitz:2011gh}. The inclusion of either quantum corrections or higher
curvature terms can remedy the situation by making the geometry flow
to AdS$_{2}\times\mathbb{R}^{d}$ in the deep IR \cite{Harrison:2012vy,Bhattacharya:2012zu,Knodel:2013fua,Barisch-Dick:2013xga}. However,
the extensive ground state entropy of AdS$_{2}\times\mathbb{R}^{d}$
has led to the idea that the true IR endpoint of the flow may be a
different geometry, such as a striped phase \cite{Donos:2011bh,Donos:2011qt,Cremonini:2012ir,Cremonini:2013epa,Hartnoll:2014gaa}, a lattice \cite{Bao:2013fda}, or a Bianchi-class geometry 
\cite{Iizuka:2012iv,Kachru:2013voa}. Even though the ultimate fate of the theory in the deep
IR is still unclear, it appears that there is a variety of possible
candidate groundstates, and thus a variety of different near-horizon
geometries. From a boundary perspective, the geometric resolution
of the horizon can be thought of as introducing a low-energy regulator.
However, in the low energy, large momentum limit,
the holographic Green's function becomes independent of the geometry
in the deep IR, up to exponentially small corrections. In that sense,
the field theory seems to care little about the exact mechanism
that resolves the Lifshitz horizon. In particular, we may speculate
that horizon features at small transverse length scales are practically
invisible at the boundary. It would be interesting to confirm this
for the case of striped phases/lattices, or a non-translationally
invariant Bianchi geometry at the horizon.

Along the same lines, it would be interesting to understand how the
tidal singularity at the horizon is reflected in field theory two-point
functions. We can try to answer this question using what we learned
about the relation between tunneling barriers and spectral functions:
Consider a bulk state with fixed momentum $|\vec{k}|$, and send $\omega\rightarrow0$.
For a black hole geometry, this corresponds to a probe falling towards
the horizon. Since the spectral function is proportional to $e^{-\alpha}$,
with $\alpha\sim|\vec{k}|^{2}/\omega^{2/z}$, it is in fact
not analytic at $\omega=0$. Although this behavior is in principle
allowed, it is certainly a peculiar feature. Moreover, as we saw in
section \ref{sec:rgflows}, the non-analyticity is absent in the case of the nonsingular
Lifshitz to AdS$_{2}\times\mathbb{R}^{d}$ flows - the spectral
function only scales as $\chi\sim e^{-|\vec{k}|}$. Thus one may speculate
that the tidal singularity in Lifshitz spacetime is mirrored in a
non-analyticity of the holographic spectral function. We hope that a further analysis of this connection will provide an answer to the interesting question ``What is the
holographic dual of a tidal singularity?''

While it may seem that any spacetime with non-relativistic scaling
in the bulk exhibits an effective tunneling barrier with height controlled by $|\vec{k}|$, our results for
$z=2$ Schr\"odinger space show that this is not the case. Here the Green's function is in fact similar to the AdS case, instead of the $z=2$ Lifshitz case, as one might have naively
suspected. A possible explanation is the appearance of an additional special conformal generator in the algebra $\mathrm{Schr}_{z=2}$, which is absent for any other $z$ \cite{Son:2008ye,Balasubramanian:2008dm,Blau:2009gd}. 
We hope to shed more light on the precise connection between bulk symmetries
and features of the Green's function in future work.

The insight that tunneling barriers correspond to exponentially suppressed
information at the boundary is not a new one. In particular, similar
observations have been made in the context of finite temperature theories.
Introducing a finite $T$ may result in an effective tunneling barrier
in the equations of motion, and as a result there are modes that are
exponentially suppressed at the boundary \cite{Son:2002sd,Horowitz:2009ij,Basu:2009xf}.
A possible future direction would be to explore the case of Lifshitz
spacetime with $T\neq0$, and study the interplay between the tunneling
barriers discussed here and the effects of a nonzero temperature.

Our results could also have direct relevance for applications
to strongly coupled condensed matter systems. In general, the spectral
function can be thought of as a measure for the density of states
in a physical system. As a result, it can be used to calculate transport
coefficients, such as thermal and electrical conductivities. An interesting
open question is whether or not the universality of the $\omega\rightarrow0$,
$|\vec{k}|\rightarrow\infty$ limit of spectral functions in Lifshitz-like
geometries can be used to extract universal predictions for conductivities
or other physical quantities in the corresponding field theories.
Such an analysis would require going beyond the case of scalar fields considered
here, to study the effect of tunneling barriers on spin-1/2 and spin-1
probes, as was done in
\cite{Hartnoll:2011dm,Iizuka:2011hg,Alishahiha:2012nm,Hartnoll:2012rj,Hartnoll:2012wm,Gursoy:2012ie}. 

By now it is clear that there exist a multitude of interesting spacetimes
that serve as candidates for holographic duals of strongly-coupled
condensed matter systems. These models have survived various nontrivial
checks and revealed many striking new features of strongly correlated
systems. However, what is still lacking at this point is a model-independent,
testable prediction of AdS/CMT, perhaps similar in nature to the celebrated
$\eta/s$ hydrodynamic bound \cite{Kovtun:2004de}. We believe that the connection
between non-relativistic scaling symmetries, tunneling phenomena and
boundary Green's functions outlined in this paper is one among many
possible paths that could ultimately lead to a universal prediction
of AdS/CMT.

\section*{Acknowledgments}
The authors would like to thank Jan de Boer, Sera Cremonini, Nabil Iqbal, John McGreevy, Niels Obers, James Sully and Kai Sun for fruitful discussions.
This work was supported in part by the US Department of Energy under grant DE-SC0007859.

\appendix

\section{Error analysis for the WKB approximation}
\label{sec:WKB}
Here we give a brief discussion of the
accuracy of the WKB approximation. The wavefunction \eqref{eq:WKBansatz}
is only the leading order approximation to the exact result. We can
parametrize a finite error in our approximation by writing
\begin{equation}
\phi_{3/4}=\sqrt{\nu}\left(U-\omega^{2}\right)^{-\frac{1}{4}}\left(1+\delta\right)e^{\pm S\left(\rho,\rho_{0}\right)},\qquad\delta\ll1.
\end{equation}
This error propagates to the matching coefficients in ${\cal M^{\prime}}$
in the following way:
\begin{equation}
{\cal M}^{\prime}=\left(\begin{array}{cc}
\left(1+{\cal O}\left(\delta\right)\right)\epsilon^{\nu}e^{S\left(\epsilon,\rho_{0}\right)} & {\cal O}\left(\delta\right)\epsilon^{\nu}e^{-S\left(\epsilon,\rho_{0}\right)}\\
{\cal O}\left(\delta\right)\epsilon^{-\nu}e^{S\left(\epsilon,\rho_{0}\right)} & \left(1+{\cal O}\left(\delta\right)\right)\epsilon^{-\nu}e^{-S\left(\epsilon,\rho_{0}\right)}
\end{array}\right).\label{eq:Mprimeerrors}
\end{equation}
While ${\cal M}_{AD}^{\prime}\rightarrow0$ for $\epsilon\rightarrow0$,
${\cal M^{\prime}}_{BC}$ actually blows up in this limit. This means
that we have no theoretical control over this coefficient, and results
containing ${\cal M^{\prime}}_{BC}$ cannot be trusted. There is
a simple explanation for this problem: We perform the matching at
$\epsilon\rightarrow0$, where the $A$-mode generically blows up, but
the $B$-mode goes to zero. For a generic solution with $A,B\neq0$,
we can then take an arbitrary finite amount of $B$ and ``hide'' it
under the non-normalizable mode $A$ by taking $B\rightarrow B-\delta B$
and $A\rightarrow A+\delta B$ . The relative error we make by doing
so will always be shrunk to zero near the boundary. This means that
generically, we cannot trust the WKB-calculation of $B$. However,
any result that does not contain the ``mixing''-term ${\cal M^{\prime}}_{BC}$
can still be calculated accurately. For example, we can calculate
$B$ for a normalizable wavefunction, where $A=0$. In this case,
we need to choose $a=-ib$ and we obtain
\begin{equation}
\left(\begin{array}{c}
A\\
B
\end{array}\right)=\left(\begin{array}{c}
0\\
{\cal M}_{BD}^{\prime}e^{-i\frac{\pi}{4}}b
\end{array}\right).
\end{equation}

Since ${\cal M^{\prime}}_{BC}$ automatically shows up in the expression
for the Green's function \eqref{eq:Ggen}, one might expect that we
cannot trust this result. However, once we plug in \eqref{eq:Mprimeerrors},
we see that
\begin{equation}
G_{\mathrm{WKB}}(\omega,\vec{k})=K\left(\frac{{\cal M^{\prime}}_{BC}}{{\cal M^{\prime}}_{AC}}+\frac{i}{2}\frac{{\cal M^{\prime}}_{BD}}{{\cal M^{\prime}}_{AC}}\frac{1-i\frac{b}{a}}{1+i\frac{b}{a}}\right),
\end{equation}
so the problematic term only appears in the real part of the Green's
function. This means that while we cannot trust WKB for $\mathrm{Re}\,G(\omega,\vec k)$,
we can still get accurate results for the imaginary part, up to an $O(\delta)$-error. In particular, one can check that ${\cal M}_{BC}^{\prime}\sim\epsilon^{-2\nu}$
and ${\cal M}_{AD}^{\prime}\sim\epsilon^{2\nu}$ do not conspire with
each other to make this error divergent.
%
\bibliography{Green}


\end{document}